\documentclass[twocolumn]{aastex631}

\usepackage{amsmath}
\usepackage{empheq}
\usepackage{graphicx}
\expandafter\let\csname equation*\endcsname\relax
  \expandafter\let\csname endequation*\endcsname\relax 
\usepackage{mathrsfs}
\usepackage{textcomp}
\usepackage{enumitem}   
\usepackage{gensymb}
\usepackage{hyperref}
\usepackage{graphicx}
\usepackage[caption=false]{subfig}
\usepackage{multirow}
\usepackage{longtable}
\usepackage[flushleft]{threeparttable}
\usepackage{booktabs} % To thicken table lines
\usepackage{CJK}

\hypersetup{colorlinks, linkcolor={blue}, citecolor={blue}, urlcolor={blue}}

\definecolor{DarkOrange}{RGB}{204, 85, 0}
\definecolor{LincolnGreen}{RGB}{17, 102, 0}

\usepackage{soul}

\usepackage{xspace}

\newcommand\nicer{\textit{NICER}\xspace}

\newcommand\swift{\textit{Swift}\xspace}
\newcommand\xmm{\textit{XMM-Newton}\xspace}

\newcommand\hst{\textit{HST}\xspace}
\newcommand\swiflong{\textit{Neil Gehrels Swift Observatory}\xspace}

\newcommand\Rin{$R_{\rm in}$\xspace}
\newcommand\Rins{$R_{\rm in}^{*}$\xspace}
\newcommand\Tp{$T_{p}$\xspace}
\newcommand\Rratio{$R_{\rm out}/R_{\rm in}$\xspace}

\newcommand\Mbh{$M_{\rm BH}$\xspace}
\newcommand\ergs{erg s$^{-1}$\xspace}

\newcommand\msun{$M_\odot$\xspace}

\newcommand\hlx{HLX-1\xspace}
\newcommand\ftli{ASASSN-14li\xspace}

\newcommand\dt{$\Delta t$\xspace}
\shorttitle{Measuring the sizes of compact accretion disks}
\shortauthors{Guolo \& Mummery}

\begin{document}
\pagenumbering{arabic}

%\title{\texttt{diskSED}: a flexible disk model for self-consistent UV/optical/X-ray fitting of accreting black holes
%}

%\title{The compactness of accretion disks from self-consistent X-ray/UV/optical fitting}

\title{The size of accretion disks from self-consistent X-ray spectra and UV/optical/NIR \\  photometry fitting: applications to ASASSN-14li and HLX-1}

\author[0000-0002-5063-0751]{Muryel Guolo}
\affiliation{Bloomberg Center for Physics and Astronomy, Johns Hopkins University, 3400 N. Charles St., Baltimore, MD 21218, USA}

\author{Andrew Mummery}
\affiliation{Oxford Theoretical Physics, Beecroft Building, Clarendon Laboratory, Parks Road, Oxford, OX1 3PU, UK}

\begin{abstract}
We implement a standard thin disk model with the outer disk radius ($R_{\rm out}$) as a free parameter, integrating it into standard X-ray fitting package to enable self-consistent and simultaneous fitting of X-ray spectra and UV/optical/NIR photometry. 
We apply the model to the late-time data ($\Delta t \approx 350-1300$ days) of the tidal disruption event (TDE) ASASSN-14li. We show that at these late-times the multi-wavelength emission of the source can be fully described by a bare compact accretion disk. We obtain a black hole mass ($M_{\rm  BH}$) of $7^{+3}_{-2}\times10^{6} M_{\odot}$, consistent with host-galaxy scaling relations; and an $R_{\rm out}$ of $45 \pm 13 \, R_{\rm  g}$, consistent with the circularization radius, with possible expansion at the latest epoch. We discuss how simplistic models, such as a single-temperature blackbody fitted to either X-ray spectra or UV/optical photometry, lead to erroneous interpretations on the scale/energetics of TDE emission. We also apply the model to the soft/high state of the intermediate-mass black hole (IMBH) candidate HLX-1. The model fits the full spectral energy distribution (from X-rays to NIR) without needing an additional stellar population component. We investigate how relativistic 
effects improve our results by implementing a version of the model with full ray tracing calculations in the Kerr metric. For HLX-1, we find $M_{\rm BH} = 4^{+3}_{-1} \times 10^{4} M_{\odot}$ and $R_{\rm out} \approx {\rm few} \times 10^{3} \, R_{\rm g}$, in agreement with previous findings. The relativistic model can constrain the inclination ($i$) of HLX-1 to be $10^o \leq i \leq 70^o$.
\end{abstract}

\vspace{1em}

\section{Introduction}

Bright disk systems evolving around compact objects offer a natural observational probe of the physics of astronomical black holes and the process of accretion itself. In particular spectral fitting, where the broad band spectral energy distribution (SED) observed from a source is used to constrain the free parameters of accretion models, is a well established technique which has been used throughout the literature to, for example, constrain the spins of Galactic X-ray binaries  \citep[e.g.,][]{Li2005}. 

The vast majority of spectral fitting models of accretion disks assume that the disk has a large (or formally infinite) radial extent.  While a reasonable approximation for many accreting systems such as X-ray binaries and active galactic nuclei, which are persistent and source their material from large radii, some transient accreting systems are expected to be significantly more compact, with an outer radius potentially only an order of magnitude larger than the inner disk size. A particularly noteworthy example of an astronomical system likely to satisfy these constraints are those disks formed in the aftermath of a tidal disruption event (TDE). 

A TDE occurs when an unfortunate star is scattered onto a near-radial orbit about a supermassive black hole (SMBH) in a galactic center. When the star moves within the so-called tidal radius it will be disrupted by the SMBHs tidal force, the stellar debris from this disruption will thereafter form an accretion flow about the SMBH, powering bright transient emission  \citep[e.g.][]{Rees1988}. The tidal radius represents the relevant size scale of the forming disk and is for typical black hole and stellar parameters of the order $\sim 10$'s of Schwarzschild radii. This is significantly smaller than assumed by conventional spectral fitting models. 

The physical size of an accretion flow can however be measured, following standard spectral fitting procedures, provided that observational data which spans a wide frequency range (typically from optical/UV up to X-ray frequencies) is available. The physical reason for this is that X-ray data probes only the inner regions of the accretion flow, and therefore any optical/UV data provides tight constraints on the properties of the outer edge of the disk. It is the purpose of this paper to derive and present a spectral fitting model which can be simultaneously fit to optical/UV through X-ray data of accreting sources, with the outer disk size as a free parameter. This allows the size of astronomical disk systems to be probed from data. 

Constraints on the physical size of accretion disks form an important part of modern analysis procedures.   For example, many models of the recently discovered class of X-ray transients known as quasi-periodic eruptions \citep[hereafter QPEs;][]{Miniutti2019,Giustini2020,Arcodia2021,Arcodia2024a,nicholl2024} suggest that the large-amplitude X-ray flares observed from these systems originate from the repeated crossing of a secondary object with an accretion flow surrounding a supermassive black hole \citep{Xian2021,Linial2023, Lu2023, Franchini2023}.  In some of these works, it has been suggested that this disk will, in many systems, have been seeded by a TDE \citep{Linial2023,Kaur2023}. To test these theories, understanding the physical size of these disks is crucial. For instance, in the recent discovery of QPEs in the nearby TDE AT2019qiz \citep{nicholl2024}, the disk's outer edge was constrained using a multi-wavelength light curve fitting approach based on time-dependent disk theory \citep{Mummery2020, mummery2024fitted}. However, this method is not applicable to most other QPE sources because their (potential) originating TDEs were not observed, with GSN 069 \citep{Miniutti2019} possibly being the only exception. Therefore, a time-independent spectral fitting approach is desired. In addition, the assumption that TDEs form compact disks is one that should be tested rigorously with data.  The spectral fitting models put forward in this paper can provide such disk size constraints.

This paper is divided as follows: in \S\ref{sec:model} we derive our models, in \S\ref{sec:data} we describe our data and fitting setup, while in \S\ref{sec:14li} and \S\ref{sec:HLX} we demonstrate the application of our models to two distinct sources, the tidal disruption event ASASSN-14li and the accreting IMBH candidate HLX-1; Some notes on model usage and limitation are presented in \S\ref{sec:notes}; our conclusions are presented in \S\ref{sec:conclusion}.

We adopt a standard $\Lambda$CDM cosmology with a Hubble constant $H_0=73\,{\rm km\,s^{-1}\,Mpc^{-1}}$ \citep{Riess2022}. 
When parameters inferred from the fitting are described as a central value plus or minus some uncertainty, the central value 
represents the median of the parameter posterior, and the uncertainties correspond to the bounds that contains 68\% of the 
posterior probability. Note that this definition differs from the frequentist definition historically used in X-ray studies \citep[see][for relevant discussion]{Andrae2010,Buchner2014,Buchner2023}.

\section{The Model}\label{sec:model}

\subsection{Newtonian regime}\label{sec:diskSED}

An observer (subscript \textit{o}) at large distance $D$ from an accretion disk observes the frequency-specific flux density $F_\nu$, which is formally given by

\begin{equation}\label{eq:1}
    F_\nu(\nu_o) = \int I_{\nu}(\nu_o) \, {\rm d}\Theta_o .
\end{equation}

\noindent Here, $\nu_o$ is the photon frequency and $I_ {\nu}(\nu_o)$ the specific intensity, both measured at the location of the distant observer. The differential element of solid angle subtended by the disk contribution on the observer’s sky is ${\rm d}\Theta_o$. In the Newtonian limit, in which energy shifting of photons (both gravitational and Doppler) and gravitational lensing are neglected, the differential element of solid angle can be written as

\begin{eqnarray}\label{eq:2}
    {\rm d}\Theta_o = \frac{ {\rm cos} \ i}{D^2} \ R \ {\rm d} R \, {\rm d}\theta,
\end{eqnarray}

\noindent where $R$ and $\theta$ are the polar coordinates in the disk frame,  $i$ is the inclination of the disk's axis with respect to the line of sight of the observer, and $D$ is the luminosity distance. In this limit, the emitted ($\nu_e$) and observed frequencies are the same\footnote{We also neglect cosmological red-shifting in this work, which could be simply included by taking $\nu_{\it o} = \nu_{\it e}/(1 + z)$ for red-shift $z$, and multiplying the amplitude of $I_\nu$ by $1/(1+z)^3$. These correction factors will be added when fitting observations.}, such that $\nu_o=\nu_e=\nu$.

\begin{figure}
	\centering
	\includegraphics[width=1\columnwidth]{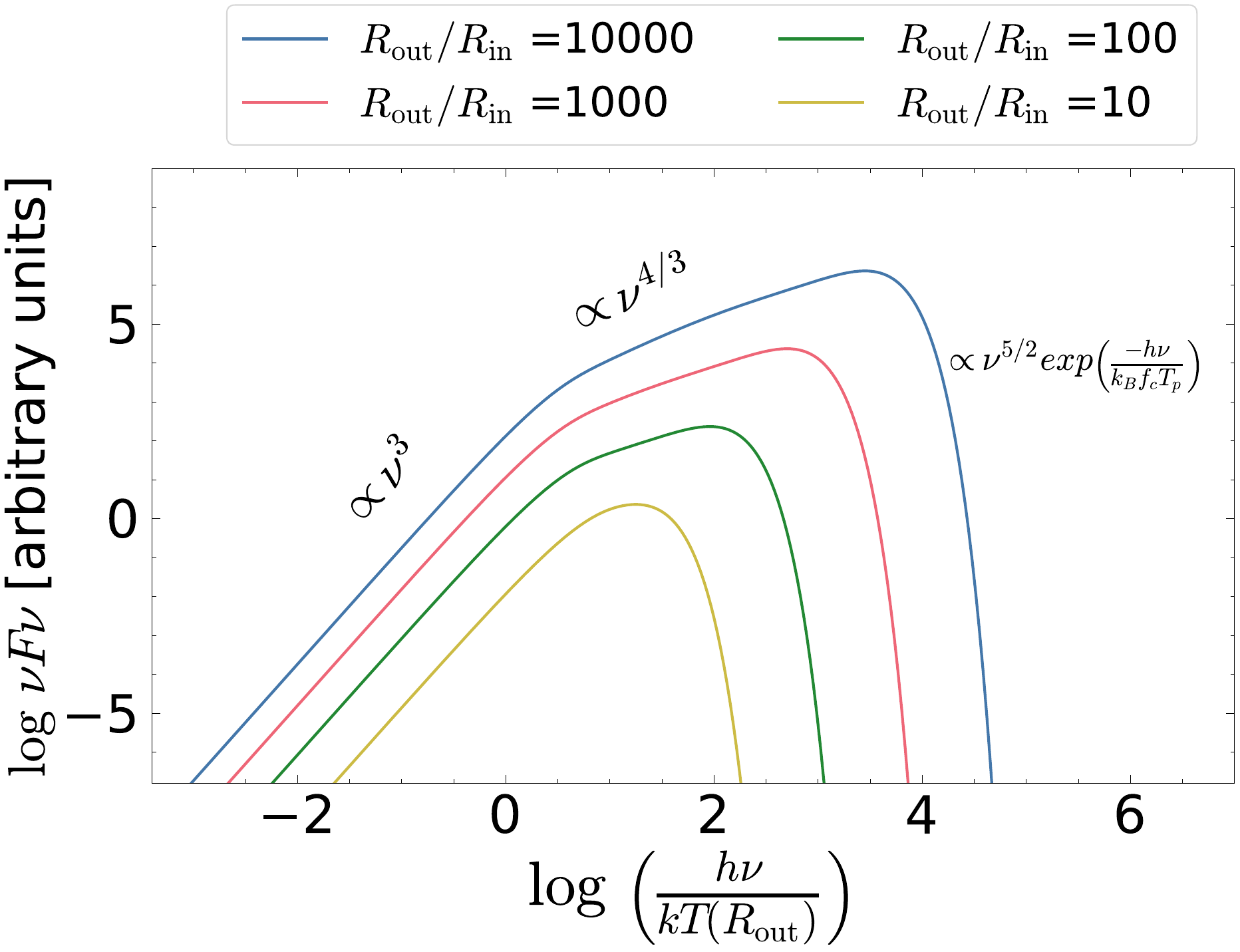}
	\caption{Broad-band spectrum shape of \texttt{diskSED} as function of the disk size ($R_{out}/R_{in}$), $x$-axis normalized by the characteristic frequency of the outer edge of the disk. $y$-axis is normalized arbitrarily for visualization purposes. The text in black shows the asymptotic shape of the broad-band spectrum in distinct frequency ranges.}

 %\cmtmum{Your last scaling on this figure is wrong! Don't worry, this fact is appreciated by basically non-one (including Frank King and Raine...). It should be $\nu^{5/2} \exp(\dots)$. Reference for this is me and Steve 2020 (14li paper). }}
    \label{fig:model_norm}
\end{figure}

The disk is assumed to be a (color-corrected) multi-temperature blackbody, each disk annulus having a temperature $T(R)$. As we shall model disk solutions at high temperatures, radiative transfer in the atmosphere of the disk from electron scattering and metal opacity effects are relevant, and here are incorporated via a simple spectral hardening factor $f_{c}$ \citep{Shimura1995}. A modified Planck function then gives the specific intensity of the locally emitted radiation

\begin{eqnarray}
    I_\nu (\nu) = f_{c}^{-4} B_\nu(\nu, f_{c} T) \equiv  \frac{2h\nu^3}{f_{c}^4c^2} \left[ {\rm exp} \left( \frac{h\nu}{k_{B}f_{c}T} \right) -1 \right]^{-1}
\end{eqnarray}

\noindent where $B_{\nu}(\nu, T)$ is the Planck function. By  integrating over the disk coordinates, the flux density observed from the surface of the disk is therefore 

\begin{eqnarray}\label{eq:disk_int_1}
    F_{\nu}(\nu) = \frac{4\pi h\nu^3 \cos\,i}{D^2c^2}\int_{R_{\rm in}}^{R_{\rm out}} \frac{R \, f_{c}^{-4} \, {\rm d}R}{  {\rm exp} \left( {h\nu}/{k_{B}f_{c}T} \right) -1 }, 
\end{eqnarray}

\noindent where $R_{\rm in}$ and $R_{\rm out}$ are, respectively, the inner and outer radius of the disk. For this implementation, we use the standard \cite{Shakura1973} temperature profile,  with the zero stress inner boundary condition. Under this assumption the radial disk temperature profile is written as

%\begin{equation}\label{eq:T_r}
%    T(r) = 2.05  \; T_p  \; r^{-3/4} \; (  1 - r^{-1/2})^{1/4}
%\end{equation}

\begin{equation}\label{eq:T_r}
    T(r) =  \left ( \frac{r_{\rm max}^3}{1 - r_{\rm max}^{-1/2}} \right )^{1/4} \; T_p  \; r^{-3/4} \; (  1 - r^{-1/2})^{1/4},
\end{equation}
in this expression $r\equiv R/R_{\rm in}$, and $r_{\rm max}= 49/36$, which is the radius where the peak temperature (\Tp) occurs, i.e., $T(r_{\rm max}) = T_{\rm p}$.  %The pre-factor $2.05$ is included so that $T_p$ corresponds to the peak physical temperature of the disk which occurs at a radius $r_{\rm max}= 49/36$. 
In the range $R_{\rm in} \ll R \leq R_{\rm out}$, the classical $T(r) \propto$ $r^{-3/4}$ profile is recovered. The color-correction factor must be kept inside the integral in Eq. \ref{eq:disk_int_1} because it is a function of the local disk temperature, and hence, the radius. In this implementation, we assume the analytical expression of $f_{c}$ given by \citet{Chiang2002}, which is calibrated on \citet{Hubeny2001} numerical simulations, and written as

\begin{equation}\label{eq:f_col}
    f_{c}(T) = f_\infty - \frac{( f_\infty -1 )[1+ {\rm exp(-\nu_b / \Delta \nu)}]}{1 + {\rm exp}[(\nu_p - \nu_b)/\Delta \nu]}, 
\end{equation}

\noindent  where $\nu_p = 2.82 k_{B} T/ h$, $f_\infty = 2.3$ and $\nu_b = \Delta \nu = 5 \times 10^{15}$ Hz (in the source frame). 

Equation \ref{eq:disk_int_1} can be expressed  in a format that is convenient for integrating into existing X-ray spectral fitting packages -- such as \texttt{XSPEC} \citep{Arnaud1996} or its \textit{Python} version \texttt{pyXspec}. Combining Eq. \ref{eq:T_r} and Eq. \ref{eq:f_col}, we define a model with three free parameters

\begin{multline}\label{eq:diskSED}
     F_{E} \left (R_{\rm in}^*, T_p, \frac{R_{\rm out}}{R_{\rm in}} \right ) =\\ \frac{4\pi E^3 R_{\rm in}^{*2}}{c^2 h^3 D^2} \int_{1}^{{R_{\rm out}}/{R_{\rm in}}} \frac{r}{f_{c}^4}  \left [ {\rm exp} \left (  \frac{E}{k_{B} f_{c} T}\right ) -1 \right ]^{-1}  {\rm d}r, 
\end{multline}
   
\noindent where $R_{\rm in}^* \equiv R_{\rm in} \;  \sqrt{{\rm cos} \, i}$. We implement this model (which we call \texttt{diskSED}) in the \texttt{Python} language, in such a way that it can be easily used in \texttt{pyXspec}\footnote{The model is publicly available at: \url{https://github.com/muryelgp/diskSED}.}.

The asymptotic form of the disk spectrum resulting from  Eq. \ref{eq:disk_int_1} is well known, and can be recovered by investigating the behavior of the integral in certain characteristic frequency limits. For frequencies $\nu \ll k_{B}T(R_{\rm out})/h$ the disk spectrum is dominated by the Rayleigh–Jeans tail of the outer disk annulus. This results in $F_{\nu} \propto \nu^2$ (or $\nu F_{\nu} \propto \nu^3$).  For $\nu \gg k_{B}f_{c} T_p/h$, the integral is dominated by the inner part of the disk and the integrated spectrum is exponential suppressed, with a characteristic functional form given by a modified-Wien tail of the hottest `effective temperature' in the disk, i.e., $f_{c}T_p$ \citep{Mummery2020}. For intermediate frequencies $k_{B}T(R_{out})/h \ll \nu \ll k_{B}f_{c}T_p/h$ the integral becomes `flat' and $\propto \nu^{1/3}$ (or $\nu F_{\nu} \propto \nu^{4/3}$); the extent of this `flat' portion of the spectrum is proportional to the size of the disk (\Rratio). This general behavior is illustrated in Fig.~\ref{fig:model_norm}.

\begin{figure}[htbp!]
	\centering
	\includegraphics[width=1\columnwidth]{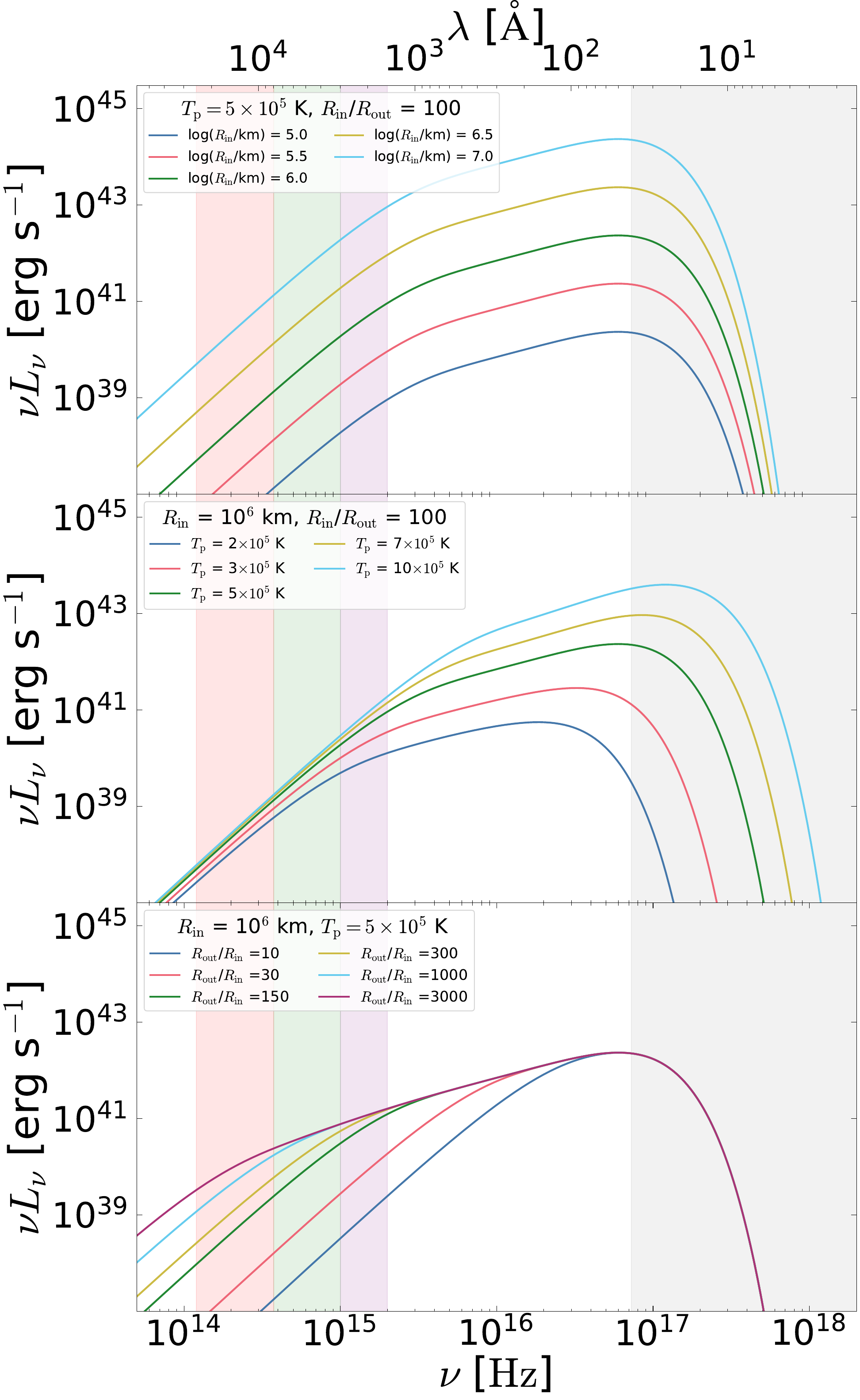}
	\caption{Broad-band spectrum of \texttt{diskSED} in physical units, as a function of each of the three free parameters: \Rins (top), \Tp (middle), and \Rratio (bottom). For reference the green spectrum is the same in the three panels. Grey regions represent the typical band for X-ray instruments (0.3-10 keV), while purple, green and red regions are, respectively, the ultraviolet, optical and near infrared bands. }
    \label{fig:model_pars}
\end{figure}

For the characteristic temperature range $10^5 \; {\rm K} \lesssim T_p \lesssim 10^6 \; {\rm K} $ the inner portion of the disk 
should produce emission which reaches into the soft X-ray band, while the outer parts of the disk are cooler and so will be detected in the low energies typical of UV/optical/IR filters. These values of $T_p$ are of interest because they are expected to be the characteristic inner disk
temperatures of disks accreting at moderate Eddington fractions around black holes with masses in the $10^3$\msun $\lesssim M_{BH} \lesssim
10^7 $\msun range. In Fig.~\ref{fig:model_pars}, we illustrate how the model's broad-band spectral energy
distribution (SED) varies in physical units depending on each of the three 
parameters in the ranges of interest.  It is important to note that in this parameter space, once the soft X-ray observations constrain the properties of the inner parts of the disk, the shape of UV/optical/IR emission is entirely controlled by the ratio \Rratio, as shown by the bottom panel of Fig.~\ref{fig:model_pars}.

The radius of the innermost stable circular orbit ($R_{\rm ISCO}$), which in our model is the inner edge of the disk and can be written as 

\begin{equation}
    R_{\rm ISCO} = \gamma(a) \frac{G M_{\rm BH}}{c^2}, 
\end{equation}

\noindent  where $\gamma(a)$ is the ISCO location in gravitational radii, which is a function of the spin,  and is written as \citep[][]{Bardeen72},

\begin{equation}\label{eq:gamma}
     \gamma(a) = 3 + Z_2 \mp \sqrt{ (3 - Z_1) (3 + Z-1 + 2Z_2)}
\end{equation}

\noindent where

\begin{equation}
     Z_1 \equiv 1 + (1 - a^2) \left [  (1+a)^{1/3} + (1-a)^{1/3}\right ]
\end{equation}
   
 \begin{equation}
     Z_2 \equiv \sqrt{3a^2 + Z_1^2}
\end{equation}

\noindent such that $\gamma(-1) = 9$ , $\gamma(0) = 6$ and $\gamma(1) = 1$. 
Consequently, once \Rins is inferred from observation it can be used to infer
\Mbh, by identifying \Rin with the ISCO, under assumptions on the inclination and spin, using

\begin{equation}\label{eq:M_BH}
    M_{\rm BH} = \frac{R_{\rm in}^* c^2 }{\gamma(a) G \sqrt{{\rm cos}\, i}}.
\end{equation}

\subsection{Fully Relativistic regime}\label{sec:model_kerrSED}

In the Kerr metric, photons do not travel in straight lines due to gravitational lensing effects, while the energy of the photons change over the course of their trajectory owing to the combined effects of kinematic and gravitational energy shifts. As a result, the relation in Eq. \ref{eq:2} is invalid, and the emitted ($\nu_e$) and observed ($\nu_o$) frequencies for a distant observer differ. The observed emission can still be expressed in a form similar to equation \ref{eq:disk_int_1}, however, since $I_{\nu}/\nu^3$ is a relativistic invariant \citep[e.g.,][]{MTW}. Utilizing this invariant, the observer-frame emission can be written

\begin{equation}
    F_{\nu}(\nu_o) = \int g^3 I_{\nu}(\nu_o/g)\, {\rm d}\Theta_o
\end{equation}
where we define the photon energy shift factor \( g \) as the ratio of observed to emitted local rest frame frequency, which is given by:

\begin{equation}\label{eq:f_g}
g(r,\phi) \equiv \frac{\nu_{{\it o} }}{\nu_{\it e}} = \frac{p_\mu U^\mu\ ({\rm O})}{p_\lambda U^\lambda\ ({\rm E})} = \frac{1}{U^{0}} \left[ 1 + \frac{p_{\phi}}{p_0} \Omega \right]^{-1},
\end{equation}

\noindent where (O) and (E) refer to quantities evaluated in the frame of the observer and emitter, respectively. 
The quantities $U^0$ and $\Omega$ are the time-like component of the disk fluid's 4-velocity, and the rate of rotation of the disk fluid, respectively. These two quantities depend on the spin $a$ and radius $r$, and are given in standard texts \citep[e.g.,][]{MTW}.
The covariant quantities $p_\phi$ and $p_0$ (on the far right) correspond to the angular momentum and energy of the emitted photon. These are constants of motion for a photon propagating through the Kerr metric.

In this case, the differential solid angle is written more generally as:

\begin{eqnarray}\label{eq:12}
    {\rm d}\Theta_o = \frac{{\rm d}b_x \, {\rm d}b_y}{D^2},
\end{eqnarray}

\noindent where \( b_x \) and \( b_y \) are photon impact parameters at infinity \citep[in effect cartesian coordinates describing the telescopes ``camera'',][]{Li2005}. Accounting again for the color-correction factor (Eq. \ref{eq:f_col}), the observed flux from Eq. \ref{eq:2} can be written in the general form for the Kerr metric as:

\begin{equation}\label{eq:flux_kerr}
    F_\nu(\nu_{\rm o}) = {1\over D^2} \iint_{\cal S} {g^3 f_{c}^{-4} B_\nu (\nu_{\rm o} /g , f_{c} T)}\,  {\text{d}b_x \text{d} b_y},
\end{equation}

\noindent where ${\cal S}$ is the disk surface define by an inner (\Rin) and an outer ($R_{\rm out}$) radius. The same $T(r)$ dependency as in Eq. \ref{eq:T_r} is assumed (with \Rin assumed to be the ISCO radius). 
In Eq. \ref{eq:flux_kerr} the observed spectrum does not depend only on the parameters of the disk ($R_{\rm in}$, $T_p$, and $R_{\rm out}$), but also on $g$, which for those photons emitted from the inner regions of the disk is a strong function of the black hole spin ($a$) and the inclination ($i$) of the disk with respect to the observer. However, except for special viewing geometries, the covariant quantities $p_\phi$ and $p_0$ in Eq. \ref{eq:f_g} must in general be found by numerical ray tracing calculations. In this implementation we employ the numerical ray tracing algorithm as described in \citet[][particularly their section 2.3.2]{mummery2024fitted}, the algorithm numerically calculates the ~$p_\phi$/$p_0$ ratios and traces back the rays of each grid-element of the observer plane back to the disk by solving the null-geodesics of the Kerr metric, allowing $g$ and hence the integral Eq. \ref{eq:flux_kerr} to be computed \textit{on the fly}; the reader is refereed to \citet{mummery2024fitted} for a more detailed understanding of the $g$ computations. 

\begin{figure}
	\centering
	\includegraphics[width=1.0\columnwidth]{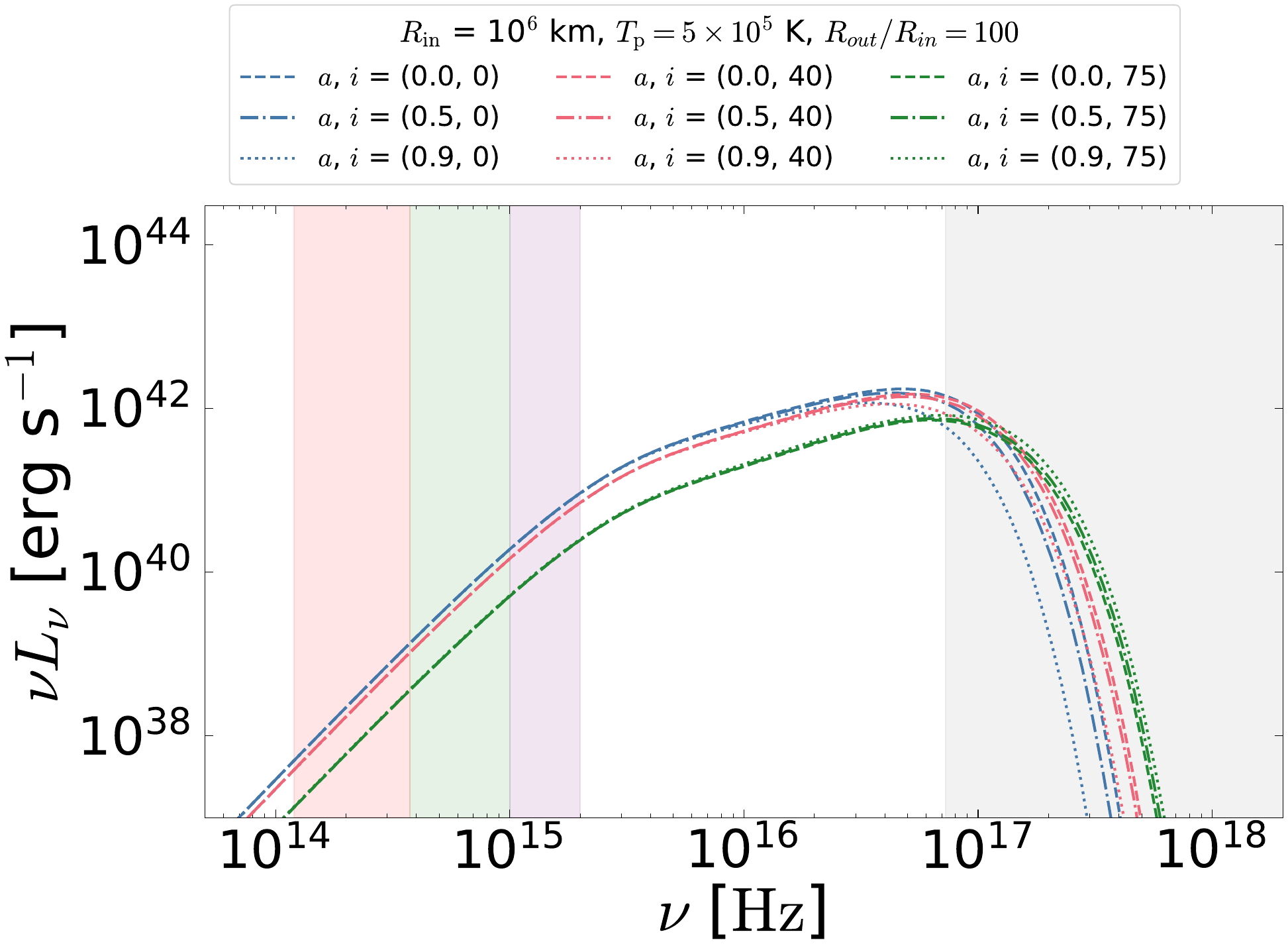}
	\caption{Broad-band spectrum of \texttt{kerrSED} as function of spin ($a$) and inclination ($i$), for fixed values of the remaining three parameters (\Rin, \Tp, and \Rratio). See text for qualitative description of their effects. Colored vertical bands are the same as in Fig.~\ref{fig:model_pars}.}
    \label{fig:model_pars_kerr}
\end{figure}

In summary, equations \ref{eq:flux_kerr}, \ref{eq:f_g}, \ref{eq:f_col} and \ref{eq:T_r} define a 5 free parameter model ($R_{\rm in}$, $T_p$, $R_{\rm out}$, $a$ and $i$), which describes a standard thin disk with vanishing stress in the innermost region including all relativistic effects of the photon propagation in the Kerr metric. 
We implement the model into \texttt{pyXspec}, which we will call \texttt{kerrSED}. The dependencies of the parameters $R_{\rm in}$, $T_p$ and $R_{\rm out}$ for \texttt{kerrSED} are similar to the ones shown in Fig.~\ref{fig:model_pars} for \texttt{diskSED}, the dependencies on $a$ and $i$, for fixed values of the other parameters, are shown Fig.~\ref{fig:model_pars_kerr}. 

Generally speaking, the effects of spin and inclination on the emergent disk spectrum can be understood physically in the following way. At  frequencies substantially lower than the peak temperature of the disk, where the emission is dominated by the detection of photons which primarily originate from the outer regions of the disk, the spin has minimal effect and the amplitude of the spectrum simply scales like $\cos i$ as in the Newtonian limit. At higher frequencies, a face-on disk generally decreases in flux for increasing spin, as the ISCO moves in towards the event horizon and more of the photons emitted from the hottest disk regions suffer larger gravitational red-shifting. On the other hand, at higher inclinations Doppler boosting can dominate, and higher spins (with faster moving inner regions) produce the largest high-energy flux.

In \texttt{kerrSED}, the normalization free parameter is \Rin instead of \Rins ($=R_{\rm in}\,\sqrt{{\rm cos} \, i}$), because in the relativistic case, the inclination ($i$) can be marginalized over during the fitting process. Consequently, the black hole mass is recovered from the values (or probability distributions) of \Rin and $a$, as:

\begin{equation}\label{eq:M_BH_kerr}
    M_{BH} = \frac{ R_{\rm in} c^2  }{\gamma(a) G}.
\end{equation}

\subsection{Comparison to other \texttt{XSPEC} models}

In this section we briefly compare \texttt{diskSED}  and \texttt{kerrSED} to other \texttt{XSPEC} models commonly used in the literature to fit X-ray spectra of sources similar to those explored in the next sections. 

The model \texttt{bbody} (or \texttt{bbobyrad}) is the simplest thermal-like model, consisting of a single-temperature blackbody. It assumes a spherical emitting geometry  from which an emission ``radius'' can be inferred. This model is not a disk model, so its best-fit parameters should not be physically interpreted when fitting an X-ray spectrum which is believed to be produced by an accretion flow. However, it can still be useful for converting counts to fluxes in non-detection X-ray observations or measuring X-ray fluxes for very low signal-to-noise spectra.

\texttt{diskbb} \citep{Mitsuda1984} is a widely used disk model. However, contrary to what is commonly assumed, \texttt{diskbb} is not an implementation of the Shakura-Sunyaev standard disk model, as it assumes $T(r) \propto r^{-3/4}$ throughout the entire disk, which is inconsistent with a zero-stress (or indeed finite-stress) inner boundary condition. \texttt{diskbb} lacks color correction, resulting in inner temperatures always being higher than the peak temperature of a more realistic disk. It also does not have a Rayleigh-Jeans tail, following $\nu F_{\nu} \propto \nu^{4/3}$ for arbitrarily large $\nu$ due to the modeling assumption of $R_{\rm out}/R_{\rm in} \rightarrow \infty$, making UV/optical/IR fitting unlikely to work for more compact disks like those in TDEs. \texttt{diskbb} is also inconsistent with a finite-stress inner boundary condition \citep[e.g.,][]{Agol2000}, as such condition would led to a distinct radii profile at large radii (e.g.,  $T(r) \propto r^{-7/8}$).

\texttt{ezdiskbb} \citep{Zimmerman2005} corrects the temperature profile of \texttt{diskbb}, assuming Eq. \ref{eq:T_r}, and is therefore consistent with a zero-stress inner boundary condition. All other properties are the same as in \texttt{diskbb} (including, importantly for our purposes, the lack of a finite disk outer edge).

\texttt{kerrbb} \citep{Li2005} includes all relevant  relativistic optics effects, and a temperature-independent color correction factor. However, like the models above, $R_{\rm out}$ is not a free parameter (it is fixed at $R_{\rm out} = 10^6 R_g$), making simultaneous X-ray and optical/UV fitting unfeasible. 

\texttt{Optxagnf} \citep{Done2012} is a standard thin disk model with a zero-stress inner boundary condition and color correction factor, which neglects photon energy shifting and lensing (similar to \texttt{diskSED}). The outer radius can be a free parameter. However, the model includes many other components related to distinct Comptonization processes, which are not necessary for our purposes, as we will show in our examples.

Finally, \texttt{tdediscspec} \citep{Mummery2021} applies a Laplace expansion to Eq. \ref{eq:flux_kerr} around the hottest region in the disk, combining the effects of $i$ and $a$ into a single parameter $\gamma$. It should recover similar values to \texttt{kerrSED} for the inner disk parameters, and inferred \Mbh. Given the expansion nature of the model, it does not need to assume a temperature radial profile, but it can only fit data taken at photon energies above the peak disk temperature, and cannot therefore be used to fit X-ray/UV/optical data simultaneously.

\section{Data and Fitting Setup}\label{sec:data}
In the following sections we aim to fit the models described above simultaneously to the X-ray spectra and UV/optical/IR photometric data of two sources, which, given our current understanding of their nature, are expected to be characterized by very distinct values of model's parameters, allowing us to probe the generalist nature of the models. The sources are the tidal disruption event (TDE) ASASSN-14li \citep{Jose2014} and the off-nuclear intermediate black hole (IMBH) candidate HLX-1 \citep{Farrell2009}.

For ASASSN-14li, in the X-rays, we focus on the high signal-to-noise ratio (S/N) data from the \xmm EPIC-pn camera \citep{Struder2001}. Thirteen observations taken at times spanning from the discovery (\dt = 0) up to \dt $\sim$ 1500 days are available. The data reduction follows the procedure described in \citet{Ajay2024}, including pile-up corrections. We also gather UV/optical photometry from the UV and Optical Telescope (UVOT) onboard \swiflong, the reduction, which is described in \citet{Guolo2024},  includes the subtraction of the host galaxy component based on the best-fitted model (and uncertainty) of the host-galaxy SED from pre-transient images of various sky surveys. In this work, we focus on the UV \textit{W2}, \textit{M2}, \textit{W1}, and optical \textit{U}-band, which are all detected above the host-galaxy level throughout the entire evolution of the source.

\hlx has shown several outbursts -- reminiscent of those observed in X-ray binaries -- 
in which the source transitions from a hard/low to a soft/high state \cite[][and 
references therein]{Soria2017}, in this work, we focus on the soft/high state of the 
2010 (MJD 55300-55700) outburst. We reduce data from X-ray Telescope (XRT) on board of 
\swift, the count rate light curve was produced using the Swift UK online tools 
\citep{Evans2009}, binned to have a S/N $\geq 3$ per bin. \textit{Hubble Space 
Telescope} (\hst) photometry is available at the soft/high state of the 2010 outburst, 
we use the values obtained by \citet{Soria2017}, as listed in their Table 1, from the 
filters {\it F140LP}, {\it F300X}, {\it F390W}, {\it F555W}, {\it F775W}, {\it F160W},
which cover wavelengths from the Far UV ($\sim 1530\;{\rm \AA}$) to the NIR ($\sim 
15370\;{\rm \AA}$).

Broad-band spectral energy distribution fitting (X-ray spectra and UV/optical/IR) is performed with the Bayesian X-ray Analysis software (BXA) version 4.0.7 \citep{Buchner2014}, which connects the nested sampling algorithm \texttt{UltraNest} \citep{Buchner2019} with the fitting environment \texttt{PyXspec}. Given the parameter inference is performed in a Bayesian framework, a probability distribution function is recovered for each parameter, which is essential for reliable uncertainty propagation on secondary parameters that can be inferred from one or more of the model's free parameters (e.g., Eq. \ref{eq:M_BH}, Eq. \ref{eq:M_BH_kerr} and Eq. \ref{eq:Rcirc}). The UV/optical/IR data were added (with no extinction correction) into \texttt{PyXspec} using the ``ftflx2xsp" tool available as part of HEASoft v6.33.2 \citep{Heasarc2014}, which creates the necessary response files to be read in the fitting package. 
While X-ray spectra alone could be fitted in its native instrumental binning using Poisson statics (a.k.a Cash statistics in \texttt{XSPEC}), \texttt{XSPEC} does not allow for UV/optical/IR data to be fitted with Poisson statics, we therefore binned the X-ray spectra using the `optimal binning' scheme \citep{Kaastra2016}, also requiring that each bin had at least 10 counts, and the simultaneous X-ray/UV/optical/IR fits were then performed using Gaussian statistics (a.k.a. $\chi^2-$statistics in \texttt{XSPEC}).

Correction for dust extinction/attenuation is essential when fitting UV and optical data. The \texttt{XSPEC} native \texttt{redden} model employs the \cite{Cardelli1989} Galactic extinction law, which will be used to correct for the Milky Way line-of-sight extinction. However, this law is not appropriate for correcting intrinsic dust attenuation in general external galaxies \citep[see][for a review on dust extinction/attenuation laws]{Salim2020}. For the intrinsic attenuation modeling, we implemented a new \texttt{XSPEC} model, which we will call \texttt{reddenSF}\footnote{This model is also available publicly alongside the disk models.}, that employs the \cite{Calzetti2000} attenuation law from 2.20 $\mu m$ to 0.15 $\mu m$, and its extension down to 0.09 $\mu m$ as described in \citet{Reddy2016}. Similar to \texttt{redden}, the relative extinction between the \textit{B} and the  \textit{V} band, E(B-V), is the free parameter of the \texttt{reddenSF} model. It is essential to notice, however, that the ratio between the specific and relative extinction R$_{V}$ = A$_{V}$/E(B-V) differs between \cite{Cardelli1989} (R$_{V} = 3.1$) and \cite{Calzetti2000} (R$_V = 4.05$) laws.

\section{Applications}
\subsection{ASASSN-14li}\label{sec:14li}

TDEs are an inevitable consequence of the existence of MBHs in the nuclei of galaxies \citep{Rees1988} and are now an observational reality, with up to $\sim$ 100 candidates identified \citep[see][for observational review]{Gezari2021}. TDEs should, in principle, provide a clean laboratory for studying the real-time formation and evolution of accretion disks in MBHs \citep[e.g.,][]{Cannizzo1990}. While the first X-ray discovered TDE candidates \citep[e.g.,][]{Komossa1999} generally agreed with the original predictions, the development of optical time-domain surveys has revealed that, at early times, several of these TDE candidates \citep[e.g.,][]{Yao2023} are much brighter in the UV/optical band and, in some cases, much fainter in X-rays \citep[e.g.,][]{Guolo2024} than what is expected from a standard thin disk, contradicting the original theoretical predictions.
\begin{figure}[h!]
	\centering
	\includegraphics[width=1.0\columnwidth]{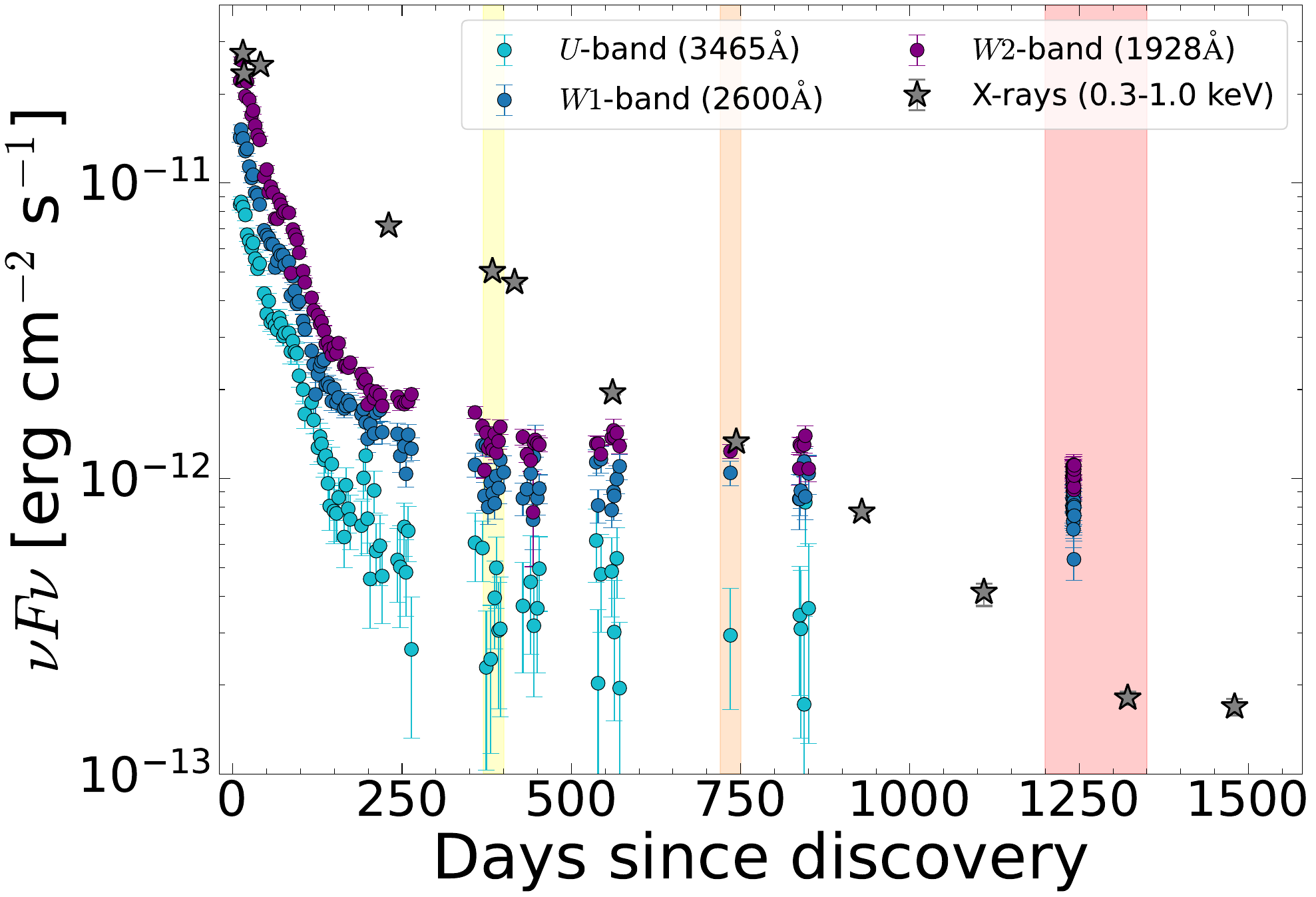}
	\caption{Multi-wavelength light curve of \ftli. Values are corrected by Galactic extinction (UV/optical) and absorption (X-rays), but not for intrinsic attenuation/absorption. Yellow, orange and red regions illustrate the three epochs analyzed in this work (E1, E2, and E3).}
    \label{fig:lc_14li}
\end{figure}
The physical origin of this discrepancy is the subject of intense debate, which can be broadly summarized as either: i) the disk formation (or circularization) is delayed, and early-time UV/optical excess emission is produced by shocks between the returning streams during the disk formation process \citep[e.g.,][]{Shiokawa2015,Ryu2023,Steinberg2024}; or ii) the disk formation is prompt, but the early-time structure of the disk differs significantly from a standard thin disk, due to the super-Eddington fallback rate, resulting in a geometrically thick disk covered by an optically thick wind/envelope/torus \citep[e.g.,][]{Metzger2016,Roth2016,Dai2018,Thomsen2022_disk_wind} that reprocesses high-energy radiation into lower energy bands.

However, as the system evolves, both scenarios seem to predict a transition to a standard thin disk phase at late times. This has been explored observationally, with multi-wavelength observations generally agreeing with such prediction during these late times \citep[e.g.,][]{Mummery2020,Guolo2024,Mummery_van_velzen2024}. In the UV/optical, this phase transition appears to be marked by a shift from a rapidly decaying light curve to a `plateau' at timescales of $\gtrsim$ 1 year after disruption \citep[e.g.,][]{vanvelzen19_late_time_UV,Mummery2024}.
\begin{figure*}[t!]
	\centering
	\includegraphics[width=0.9\textwidth]{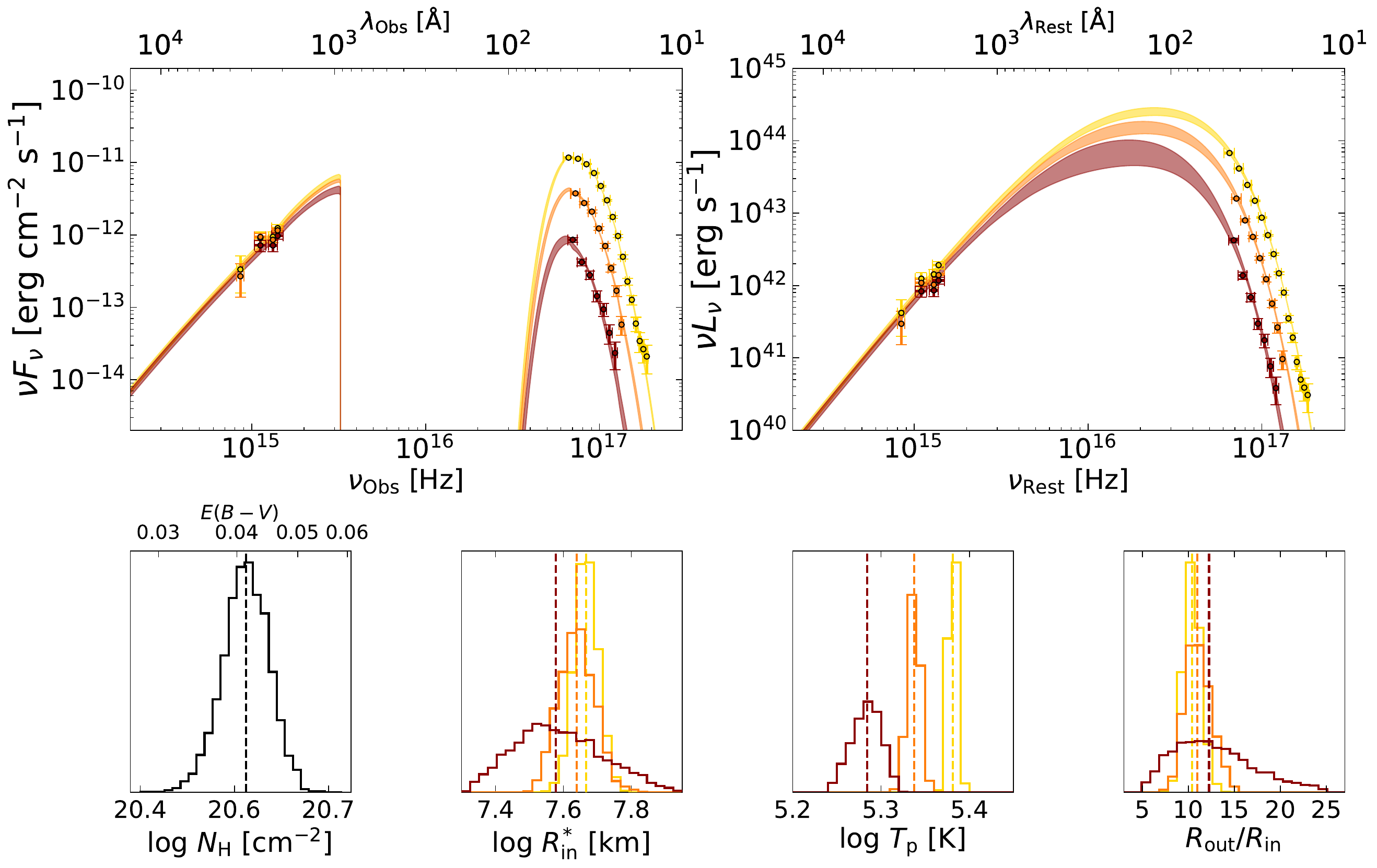}
	\caption{Results of the nested sampling fit of \texttt{diskSED} to broad-band data of three epochs of \ftli. The left upper panel shows the observed flux models (without any extinction/absorption correction) overlaid on the observed UV/optical photometry and the unfolded X-ray spectra. The right panel shows the intrinsic luminosities (with both Galactic and intrinsic extinction/absorption corrections), with the data points unfolded to the median values of the parameter posteriors. The bottom panel shows the 1D projection of marginalized posteriors of 10 free parameters. Vertical lines show the median values of the posterior distributions. The epochs correspond to $\Delta t \approx 350$ days (yellow), $\Delta t \approx 750$ days (orange), and $\Delta t \approx 1250$ days (red).}
    \label{fig:14li_sed}
\end{figure*}
The working hypothesis that the authors here wish to demonstrate is that, at these late times, the full SED from X-rays to the optical of TDEs (or for now, at least in one TDE) can be described by a simple standard thin disk and that when all the relevant physical processes are accounted for, the underlying physical parameters of the system can be inferred via self-consistent broad-band spectrum fitting. 

We selected \ftli as our study source in this paper given its low redshift and the abundance of high-quality multi-wavelength data in a long-baseline since the discovery, as shown by the light curve in Fig.~\ref{fig:lc_14li} (based on the data as described in \S\ref{sec:data}). Many studies have explored the multi-wavelength data of \ftli; however, the number of studies that apply self-consistent and physically motivated models to the X-ray and UV/optical data are more rare. %{\bf \textcolor{purple}{I agree with you, but be careful here not to piss people off...}}.

\citet{Mummery2020} developed and solved the time-dependent relativistic thin disk equations and fit to \ftli's UV/optical and integrated X-ray light curves (instead of X-ray spectra); such an approach has pros and cons. The time-dependent nature of the model allows for estimates of the parameters such as the total disk mass ($M_{\rm disk}$) and the surface density profile ($\Sigma$) of the disk, which is not possible for time-independent models (such as those described in \ref{sec:diskSED}). However, by fitting the integrated X-ray luminosity instead of the X-ray spectra,
additional information that could be obtained from the shape of X-ray spectra are lost e.g., much more precise constraints on the inner region of the disk properties. 

A distinct approach was taken by \citet{Wen2023}, the authors first fit the X-ray spectra \citep[using a time-independent slim-disk model, see ][]{Wen2022}, and then extrapolated their X-ray modeling solutions to the lower energies and compared those extrapolations to the observed UV/optical data. A more direct comparison between our work and the approach and results by \citet{Wen2023} will be discussed later but can be summarized by the fact that we will fit X-ray spectra and optical and UV photometry simultaneously. 

For our fitting, we selected three epochs (E1, E2, and E3) during the UV/optical `plateau' phase\footnote{We refer to this phase as a ``plateau'', given the slow evolution. However, it should be noted that the UV/optical flux decreases by $\sim 20\%-30\%$ from $\Delta t \sim 350$ to $\Delta t \sim 1300$ days.}, where simultaneous UVOT UV/optical photometry and \xmm X-ray spectra are available; these span from $\sim$ 380 days (E1), to $\sim$ 1250 days (E3) since discovery, the epochs are marked as yellow, orange and red vertical bands in Fig.~\ref{fig:lc_14li}. For each epoch, our total fitted model is described in \texttt{XSPEC} language as \texttt{phabs$\times$redden$\times$zashift(phabs$\times$reddenSF$\times$diskSED)}.

The Galactic X-ray neutral gas absorption is fixed to the Galactic hydrogen equivalent column density equals to $N_{H, G} = 2.0 \times 10^{20}$ cm$^{-2}$ \citep{HI4PI2016} and the Galactic extinction is given by a E(B-V)$_{G}$ = 0.022 \citep{Schlafly2011}, and modeled by \texttt{redden}. The intrinsic part of the model is shifted to the source rest frame using \texttt{zashift} with $z = 0.0206$. The three parameters of \texttt{diskSED} (\Rins, \Tp, and \Rratio) are free to vary independently in each of the three epochs. 
To jointly fit the three epochs, we start with the hypothesis that the intrinsic X-ray absorption is produced in the host galaxy and is not related to the TDE; therefore, the intrinsic $N_H$ should not vary between epochs. While it is beyond the scope of this paper to perform a Bayesian model comparison while freeing $N_H$ epoch by epoch, in a simplistic frequentist framework, allowing $N_H$ to vary in each epoch would increase the number of free parameters by $N-1$, where $N$ is the number of epochs being fitted jointly. This would require the fit with fixed $N_H$ to be of poor quality to justify the increase in free parameters. However, we will show that this is not the case.

If the X-ray absorption is caused by neutral gas in the host galaxy, then the intrinsic dust attenuation (modeled by \texttt{reddenSF}) is not completely independent but is related to the neutral gas absorption by a given gas-to-dust ratio. In normal galaxies (i.e., not long-lived active galactic nuclei), this gas-to-dust ratio should vary only mildly, depending on the galaxy's metallicity. However, the data quality here may not be sufficient to measure these deviations with statistical significance, and we therefore assume a Galactic-like gas-to-dust ratio, given by $N_H ({\rm cm}^{-2}) = 2.21 \times 10^{21} \times A_V ({\rm mag})$ \citep{Guver2009}. Thus, the model must self-consistently correct for the effects of neutral gas absorption (X-rays) and dust attenuation (UV/optical). Therefore, our final model for the joint fit of three epochs of UV/optical photometry and X-ray spectra has only $3\times3 + 1 = 10$ free parameters. Uniform priors are assumed for all the free parameters.

The results of the nested sampling fit (see \S\ref{sec:data}) are shown in Fig.~\ref{fig:14li_sed}. The bottom panel shows the 1D projection of the 10 parameter posteriors. The full posterior of all parameters is shown in Appendix \S\ref{sec:app_post}. The convergence of the sampling is clear. In the left upper panel of Fig.~\ref{fig:14li_sed}, we show the observed flux models (without extinction/absorption corrections) overlaid on the observed UV/optical photometry and the unfolded X-ray spectra. The right panel shows the intrinsic luminosities (with both Galactic and intrinsic absorption/attenuation corrections), with the data points unfolded to the median values of the parameter posteriors. The compactness of the disk is evident from the extremely short ``flat'' portion of the broad-band spectrum. 
\begin{figure}[h]
	\centering
	\includegraphics[width=1.0\columnwidth]{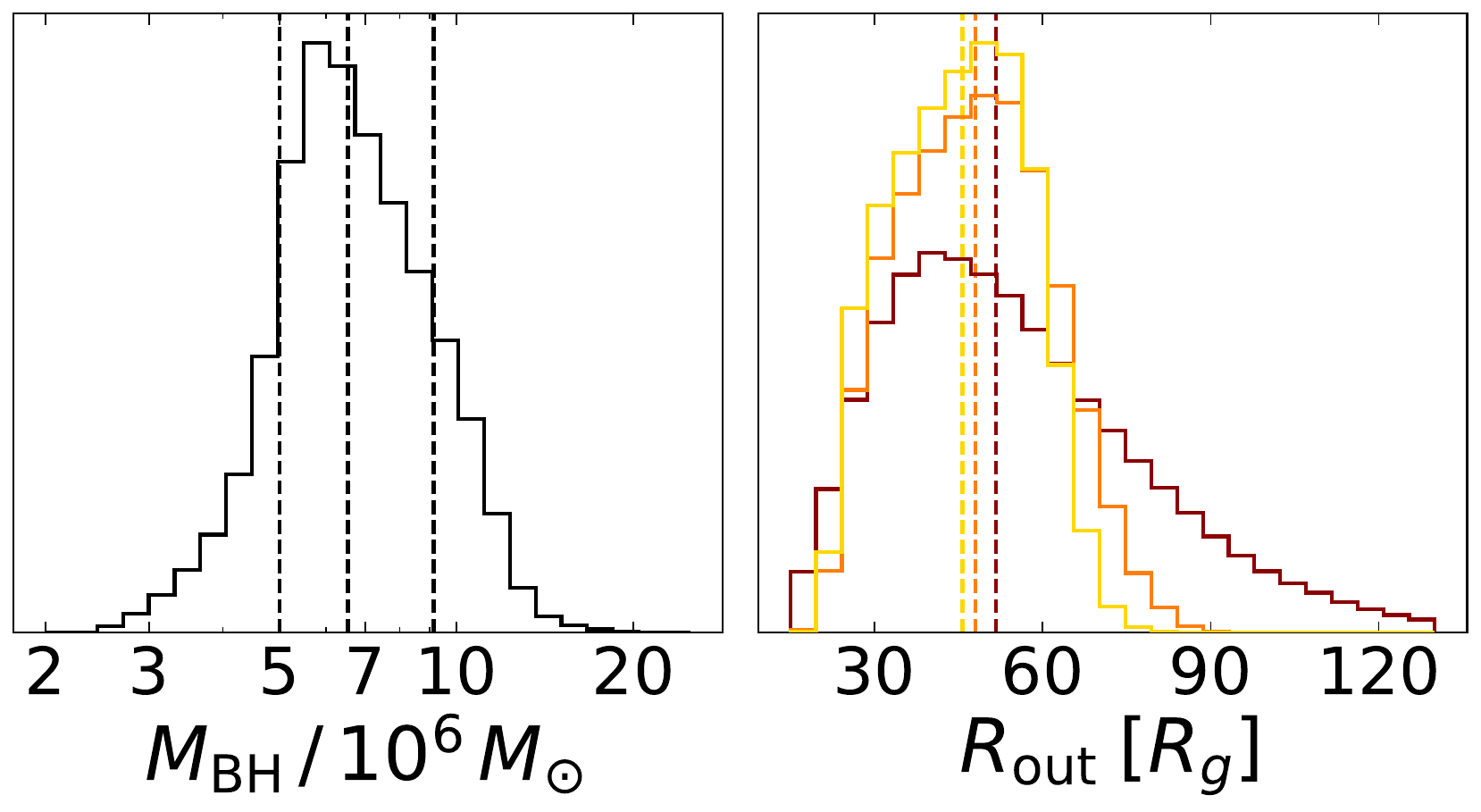}
	\caption{Probability distribution functions for \ftli's \Mbh (left panel) and outer disk radius ($R_{out}$) in gravitational radius's ($R_g$, right panel) for the three epochs. Colors scheme follows previous figures.}
    \label{fig:14li_mbh_rout}
\end{figure}

Among the disk parameters, the inner disk temperature (\Tp) shows the most significant evolution from epoch E1 to epoch E3. The posteriors for each epoch do not overlap, indicating that the cooling of the disk is recovered at high significance. This cooling is a fundamental prediction of time-dependent disk evolution theory \citep[e.g.,][]{Cannizzo1990, Mummery2020}. While this cooling had already been confirmed through analyses of X-ray spectra alone for ASASSN-14li and other TDEs \citep[e.g.,][]{Ajay2024, Wevers2024, Guolo2024, Yao2024}, it is reassuring to observe this evolution when simultaneously fitting the X-ray, UV, and optical emissions.

\begin{figure}[h]
	\centering
	\includegraphics[width=0.95\columnwidth]{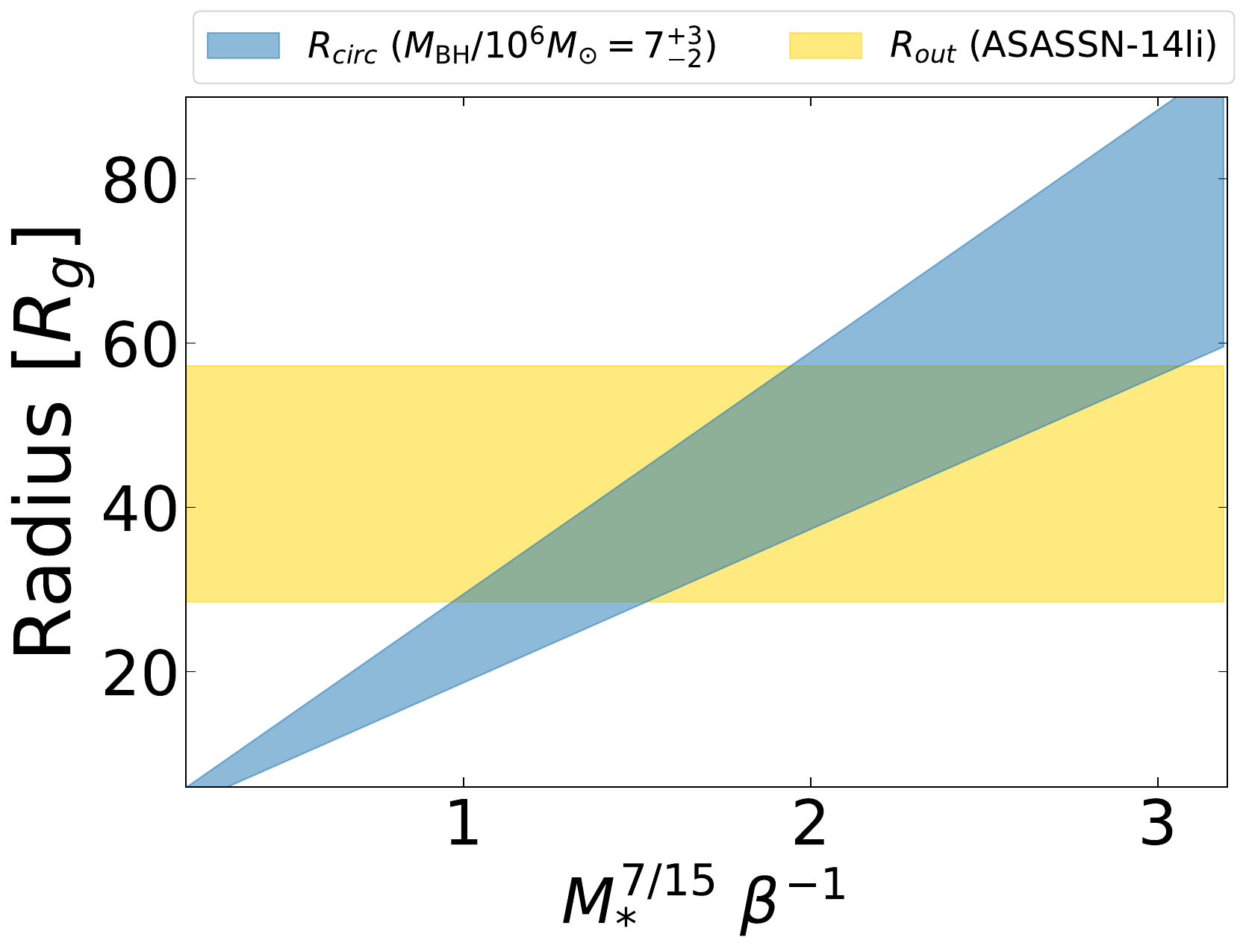}
	\caption{Comparison between the expected circularization radii ($R_{circ}$), as a function of the product $M_*^{7/15}\beta^{-1}$ (see Eq. \ref{eq:Rcirc}) and the earliest measured outer disk radii ($R_{out}$) for \ftli. Bands represent the region that contains 68\% of the probability distribution.}
    \label{fig:Rout_vs_Rcirc}
\end{figure}

The (physical size of the) inner radii of an accretion disk following a TDE should not in principle vary substantially with time, as none of the variables in Eq. \ref{eq:M_BH} should change over time\footnote{The inclination {\it i} could vary with time in the early phases if the disk is formed misaligned with the black hole spin vector \citep[see e.g.,][]{Pasham2024}; however, it should realign with the MBH spin vector at later times due to the Bardeen \& Petterson effect \citep{Stone2012}.}. Although the uncertainty on \Rins for epoch E3 is much higher than for other epochs, given the lower count-rate (hence lower S/N), the \Rins values inferred from the three epochs are consistent with each other, around ${\rm log}(R_{\rm in}^*/{\rm km}) = 7.6-7.7$.  This strengthens the case that at these late-times, the full multi-wavelength emission of \ftli is in fact described by bare disk spectrum, and that \Mbh can be inferred from \Rins using Eq. \ref{eq:M_BH}.

For the latter, in the Newtonian regime of \texttt{diskSED}, assumptions about inclinations and spin need to be made, as they cannot be marginalized over from the model. We assume a flat probability distribution of prograde spins in the $0 \leq a \leq 0.99$ range, and a flat probability distribution for ${\rm cos} \,i$, with inclinations in the range $0^\circ \leq i \leq 45^\circ$, as there are independent arguments for \ftli not being an edge-on-like system \citep[see, e.g.,][]{Dai2018, Charalampopoulos2022, Thomsen2022_disk_wind, Guolo2024}.
Combining the probability distributions of {\it i}\xspace, {\it a}, and \Rins from the 3 epochs, the probability distribution of \Mbh as shown in the left panel of Fig.~\ref{fig:14li_mbh_rout} is obtained, which can be written as $M_{\rm BH} = 7^{+3}_{-2} \times 10^{6} \ M_{\odot}$.

\begin{figure}
	\centering
	\includegraphics[width=1.0\columnwidth]{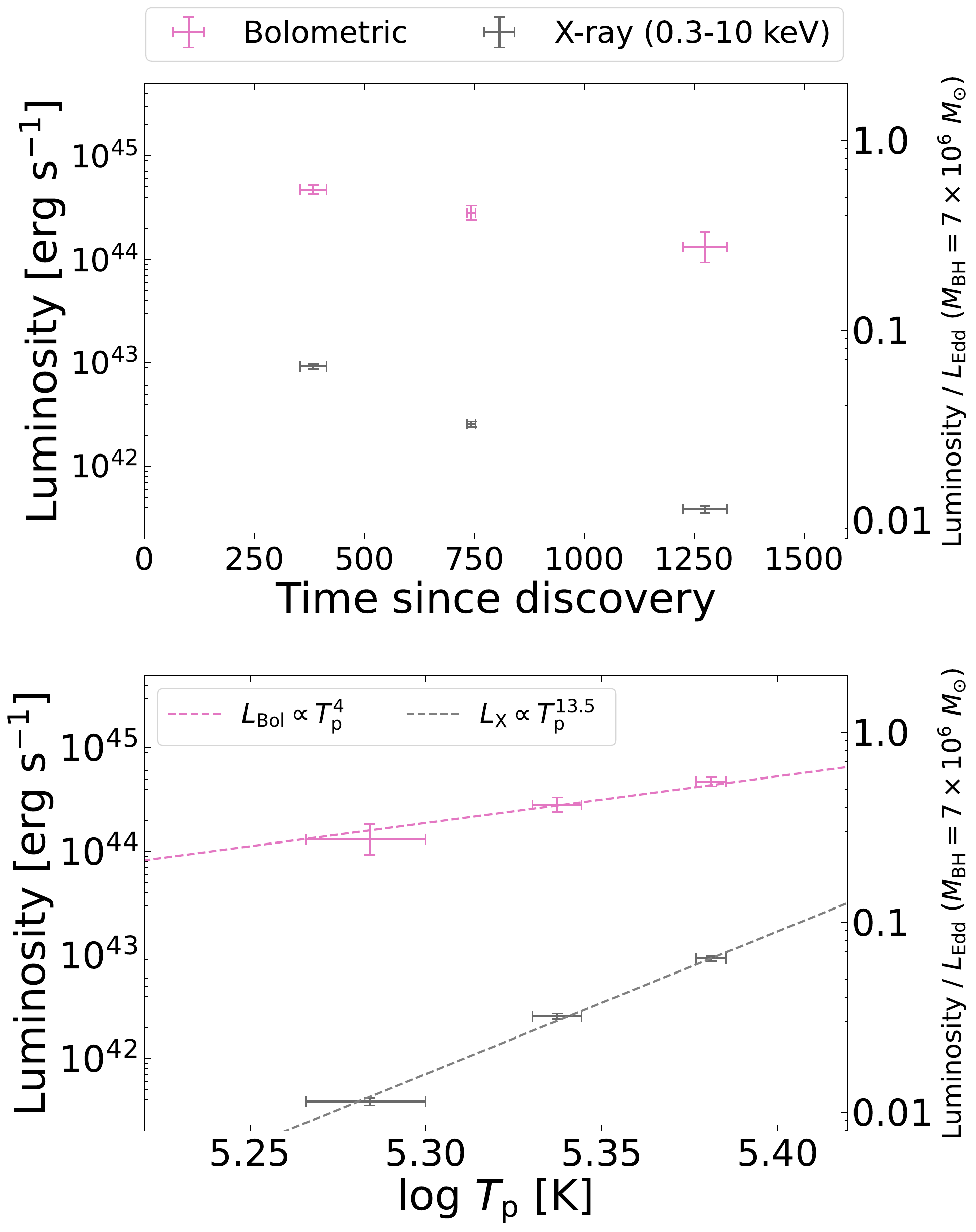}
	\caption{Evolution of Bolometric (pink) and X-ray (grey) luminosities, as a function of time since discovery (top panel) and peak disk temperature ($T_p$, bottom panel).}
    \label{fig:L_Bol_Tp}
\end{figure}
The \Mbh value obtained here is in agreement within the uncertainties to previous work using distinct X-ray continuum fitting models \citep{Wen2020,Mummery2023,Guolo2024} and with plateau-scaling by \citet{Mummery2024} (which uses only late-time optical/UV data). It is also in agreement with host-galaxy relations, as the nuclear stellar velocity dispersion of \ftli's host is $\sigma_* = 81 \pm 2$ km s$^{-1}$ \citep{Wevers2019_Mbh}, which, using \Mbh$-\sigma_*$ relations translates into values varying from ${\rm few} \times 10^{6} - {\rm few}  \times 10^{7}$ \msun depending on which of relations is applied, and given that these relations have systematic dispersions that are around ${\pm\, 0.5 \ \rm dex}$. 

At odds with our expectations -- but in agreement to \citet{Wen2023}'s findings -- is the fact that the outer radii does not increase significantly with time. Although the probability distribution of \Rratio in E3 is much more skewed to higher values, than on E1 and E2, no statistical significant claims about the expansion of the outer radii can be made with the data available, as the uncertainties on E3's parameters are larger, given the lower S/N at these very-late times. The physical reason one would expect radial expansion is that the governing temperature profile (Eq. \ref{eq:T_r}) is derived under the assumption that the locally liberated energy of the accretion process is sourced from the local redistribution of angular momentum in the flow, with angular momentum flowing outwards while the matter flows inwards. This outward flow of angular momentum should lead to disk expansion, although we note that in classical time dependent disk theory predicts a relatively weak power-law dependence with time ($R_{\rm out} \propto (t/t_{\rm evol})^{3/8}$ for the canonical \citealt{Cannizzo1990} model, where $t_{\rm evol}$ is the timescale the bolometric luminosity decays on, for example). A disk with a substantial ISCO stress on the other hand undergoes minimal radial expansion over the first phase of its expansion \cite{Mummery19b}, which for a TDE disk could be of order $\sim$ years.

Perhaps, more interesting than the \Rratio would be the value of $R_{\rm out}$ itself, however, to go from \Rratio to $R_{\rm out}$ in physical units (e.g., km) one would need to make assumption on both {\it a} and {\it i}. But, if instead we express $R_{\rm out}$ in $R_g$'s, it can be easily shown that the dependency on {\it i} cancels out, which decreases the uncertainty in derived values. By assuming the same flat distribution of spins in the $0 \leq a \leq 0.99$ range, the probability distributions for $R_{\rm out}/R_g$, as shown in the right panel of Fig.~\ref{fig:14li_mbh_rout}, are obtained. Naturally, the skewed distribution on E3 is maintained, allowing for $R_{\rm out} \leq 120 R_g$ (99\% of the posterior), but still statistically consistent with the $R_{\rm out} = 45 \pm 13 R_g$ obtained in E1. The $R_{\rm out}/R_g$ value obtained in E1, is in agreement with the values obtained by \citet{Wen2023}, which by exploring several extinction/attenuation laws with several values of E(B-V), obtained value that varied from  $10 \leq R_{\rm out}/R_g \leq 55$ ($1\sigma$ values), while our smaller uncertainties arise from the fact that our broad-band fitting was performed simultaneously and self-consistently using a fixed gas-to-dust ratio, as described above. 

The $R_{\rm out}$ value obtained from E1 is of particular interest, because it is the earliest epoch in which the size of the newly formed disk can be measured, and there are theoretical expectations for the extent of disks formed in the aftermath of TDEs. From simple conservation of angular momentum arguments, one can show that such disk should be as extended as the so call `circularization radius' ($R_{\rm circ}$), which is defined as two times the periapsis radius ($R_p$) of the disrupted star, and can be written as 

\begin{equation}
    R_{\rm circ} = \frac{2R_T}{\beta}
\end{equation}

\noindent where $R_T$ is the tidal radius, $\beta$ is the impact parameter, defined as the $R_p/R_T$ ratio. The extra factor two here originates from conservation of angular momentum as a parabolic incoming orbit is turned into a circular disk orbit. In addition, the tidal radius can be written as a function of the black hole and disrupted star properties, such that

\begin{equation}
    R_T \approx R_* \left (  \frac{M_{BH}}{M_*}\right )^{1/3}.
\end{equation}

\noindent Therefore, for a main-sequence star, where $R_* \propto M_*^{4/5}$, $R_{\rm circ}/R_g$ can be written as a function of the \Mbh, $M_*$, and $\beta$, as

\begin{equation}\label{eq:Rcirc}
    \frac{R_{\rm circ}}{R_g} (M_{BH}, M_*, \beta) \approx  \frac{2c^2R_{\odot}}{\beta G M_{\odot}} \left ( \frac{M_*}{M_{\odot}} \right)^{7/15} \left ( \frac{M_{BH}}{M_{\odot}} \right)^{-2/3},
\end{equation}

\begin{figure}[htbp!]
	\centering
	\includegraphics[width=1\columnwidth]{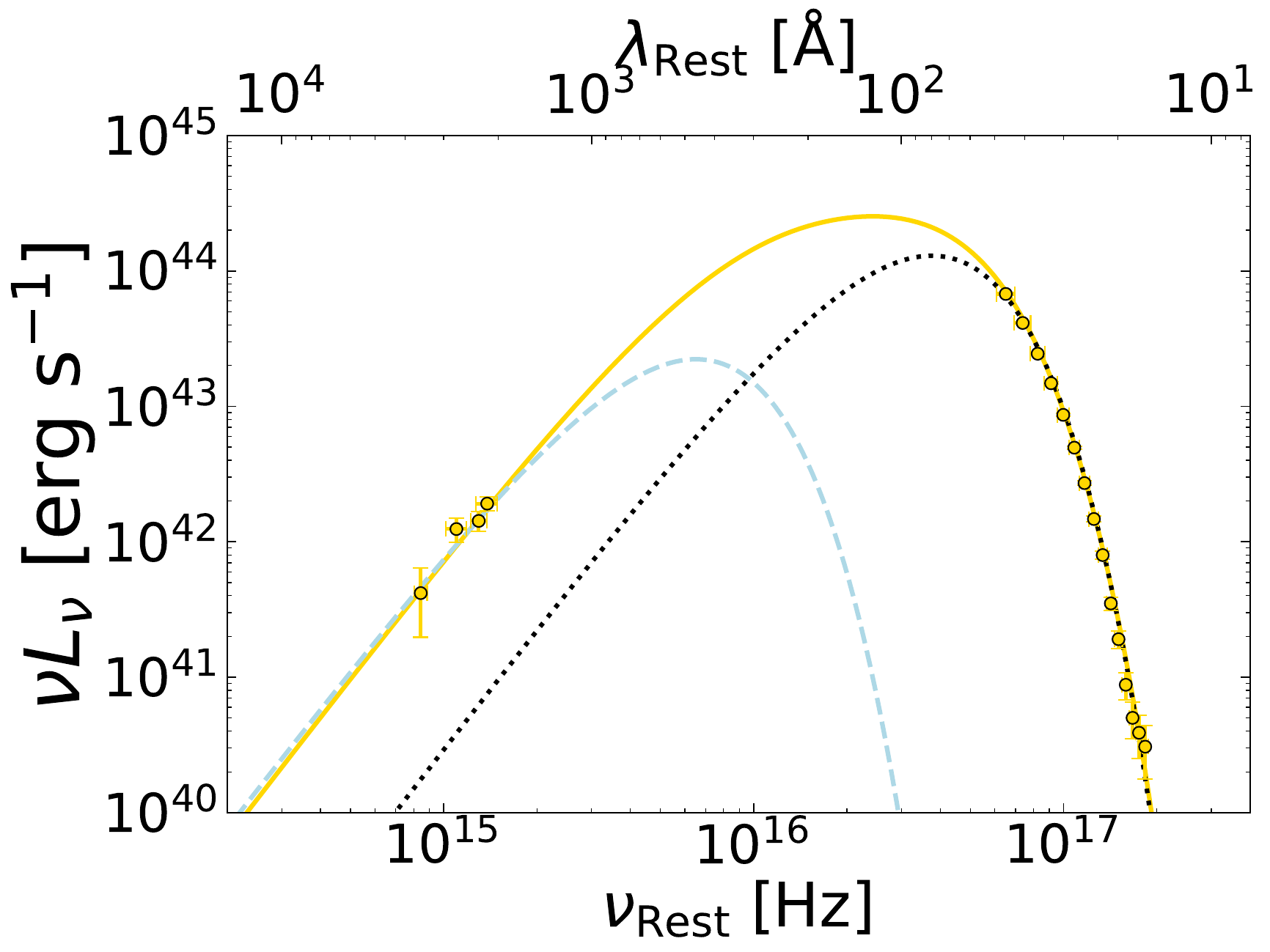}
	\caption{Comparison between a physically motivated disk model (\texttt{diskSED}) fitted simultaneously to X-ray spectra and UV/optical photometry (yellow) of \ftli, with single-temperate blackbodies fitted to either X-ray spectra (dotted black) or UV/optical photometric SED (dashed cyan). Single temperature blackbody functions will always underestimate the Bolometric luminosity, and may lead to erroneous interpretations on the scale and energetics of TDEs (see text for discussion).}
    \label{fig:BB_comp}
\end{figure}

\noindent such that for a given \Mbh value, or probability distribution (as in Fig.~\ref{fig:14li_mbh_rout}), the expected outer radii of a disk formed following a TDE depends linearly on the $\beta^{-1}M_*^{7/15}$ product, and can be compared with the value obtained from our fit of E1. In Fig.~\ref{fig:Rout_vs_Rcirc}, we show that $R_{\rm out}$ derived for \ftli is consistent with the expected $R_{circ}$ as long as $\beta^{-1}M_*^{7/15} \geq 1$, which simply requires that the mass of the disrupted star was $M_* \geq 1$ \msun. A lower initial stellar mass is possible if the disk underwent some radial expansion prior to the first observation used here, which was taken $350$ days post peak (as might be expected from the shallow $R_{\rm out} \sim t^{3/8}$ power-law predicted from time-dependent disk theory).

The bolometric luminosity ($L_{\rm Bol} \equiv \int_{0}^{\infty} L_{\nu}(\nu) {\rm d}\nu$) in most of the TDE literature is estimated using a single-band 
``bolometric-correction'' factor ($k$), such that $L_{\rm Bol} = k \times \nu L_{\nu}$, where $k$ is obtained by assuming that the model 
fitted to this narrow frequency range (e.g., the UV/optical band) describes the emission not only in this narrow band but the full frequency range. 
We have already shown that our model can self-consistently describe all the observed data available in all the wavelengths, such that our 
uncertainty on $L_{\nu}(\nu)$ is solely driven by the statistical uncertainty of the data, and the bolometric luminosity can be obtained by numerical integration, following the definition above.

In Fig~\ref{fig:L_Bol_Tp}, we show the evolution of $L_{\rm Bol}$ with time during the $\Delta t \approx 350 - 1300$ days, and with \Tp, we also show the evolution of the X-ray luminosity ($L_{\rm X} \equiv \int_{0.3\, {\rm keV} }^{10 \,{\rm keV}} L_{E}(E) {\rm d}E$), as a function of the same variables. As has already been shown by previous authors \citep[e.g.,][see their equation 91]{Mummery2020}, the X-rays not only carries just a small fraction of the total energy, but also decays much faster than $L_{\rm Bol}$ (given the simultaneous cooling of the disk and the X-ray luminosities exponential dependence on disk temperature). 
This is clearly illustrated by the fact that at $\Delta t \approx 1250$ days ($\sim$ 3.5 years after disruption), \ftli's $L_{\rm Bol}$ is still $\sim 10^{44}$ \ergs, while the X-ray luminosity has already decayed below $10^{42}$ \ergs. By simply integrating over a power-law that connects the three $L_{\rm Bol} \times t$ points in the top panel of Fig.~\ref{fig:L_Bol_Tp}, the energy emitted only during the $\Delta t \approx 350-1250$ days period is $\sim 2 \times 10^{52}$ ergs, which is mostly emitted in the Extreme UV (EUV) frequencies, and consistent with $\sim 0.1 M_{\odot}$ being accreted only in this period, in agreement with what is expected from the disruption of a star. 

One of the consequences of the cooling of the disk that shifts the radiation out of the X-ray band as the system evolves is that linear correlations between $L_{\rm X}$ and the accretion rate ($\dot{M}_{BH}$), in the form of $L_{\rm X} = \eta c^2 \dot{M}_{BH}$ (where $\eta \leq 0.1$, is the accretion efficiency) assumed by some analytical work on TDEs 
is not valid, given the nonlinearity in the relation between $L_{\rm X }$ and $L_{\rm Bol}$ and the fact that most the accretion radiation is emitted in the EUV and not in the X-rays. 

The relation $L_{\rm Bol} \propto T_{p}^{4}$ expected from a constant area disk\footnote{Given neither \Rins nor $R_{\rm out}$ had varied significantly, a constant area disk is a reasonable zeroth-order approximation for \ftli's disk.}, can approximately describe the evolution of \ftli, as shown by the bottom panel of Fig.~\ref{fig:L_Bol_Tp}. For the reasons described above, the relationship between X-ray luminosity and inner temperature is significantly steeper. Phenomenologically, this can be approximated by $L_{\rm X} \propto T_{p}^{13}$ for \ftli. However, the analytical form of this dependency is a product of power-law (describing the bolometric decay) and exponential (describing the shift of the SED as a function of temperature as it moves out of the X-ray band) functions, as detailed in section 3.6 of \citet{Mummery2020}.

In observational studies, a single temperature blackbody function is often used to model TDE emission. This approach is commonly applied to the UV/optical broad-band SED (hereafter denoted as ${\rm BB}$) and, though less frequently, also to X-ray spectra (hereafter denoted as ${\rm BB,X}$). In the X-rays, it has already been discussed extensively by \cite{Mummery2021} that although the peak ``effective temperature'' ($f_{c}T_p$) may be similar to the recovered $T_{\rm BB,X}$, the normalization (hence the recovered ``X-ray radii'', $R_{\rm BB,X}$) will have no physical meaning, and it will likely be smaller than the $R_{\rm ISCO}$.

In the UV/optical bands, the derived value for $L_{\rm BB}$ (i.e., the integrated luminosity under the single temperature blackbody assumption) is often interpreted as being equal to the bolometric luminosity. 
This interpretation is clearly incorrect in the late times of sources with observed X-ray emission (see Fig.~\ref{fig:BB_comp}).

In Fig.~\ref{fig:BB_comp}, we compare our multi-temperature disk model fitted to E1 with single temperature blackbodies fitted to either UV/optical bands or X-ray spectra. As can be clearly seen, both underestimates $L_{\rm Bol}$; even adding $L_{\rm BB}$ and $L_{\rm BB,X}$ would still underestimate $L_{\rm Bol}$. At E1 $L_{BB} \approx 3 \times 10^{43}$ \ergs, while $L_{\rm Bol} \approx 5 \times 10^{44}$ \ergs, meaning is that in this epoch/source the single-temperature underestimate the Bolometric luminosity by a factor of $\sim$ 16. The underestimation will be worsen the hotter the inner disk temperature is \citep{Mummery2020}.

However, the single temperature blackbody assumption is not a poor assumption only for source with bright X-ray emission; instead even for the early-time emission of sources where X-rays are not detected, the single temperature blackbody approximation has been shown to significantly underestimates \citep[by orders of magnitude,][]{Leloudas2019} the EUV emission needed to produce the observed He II and Bowen fluorescence emission lines commonly seen in TDEs \citep{Charalampopoulos2022}, thus also underestimating the actual bolometric luminosity.

Some studies also identify the $\int L_{\rm BB}(t){\rm d}t$ as the ``total radiated energy'', which will inevitably be less than what we obtained above using a physically motivated model and less than what is expected from the disruption of a star. Such a misidentification necessarily leads to ``missing energy'' claims.

\begin{figure}[h!]
	\centering
	\includegraphics[width=\columnwidth]{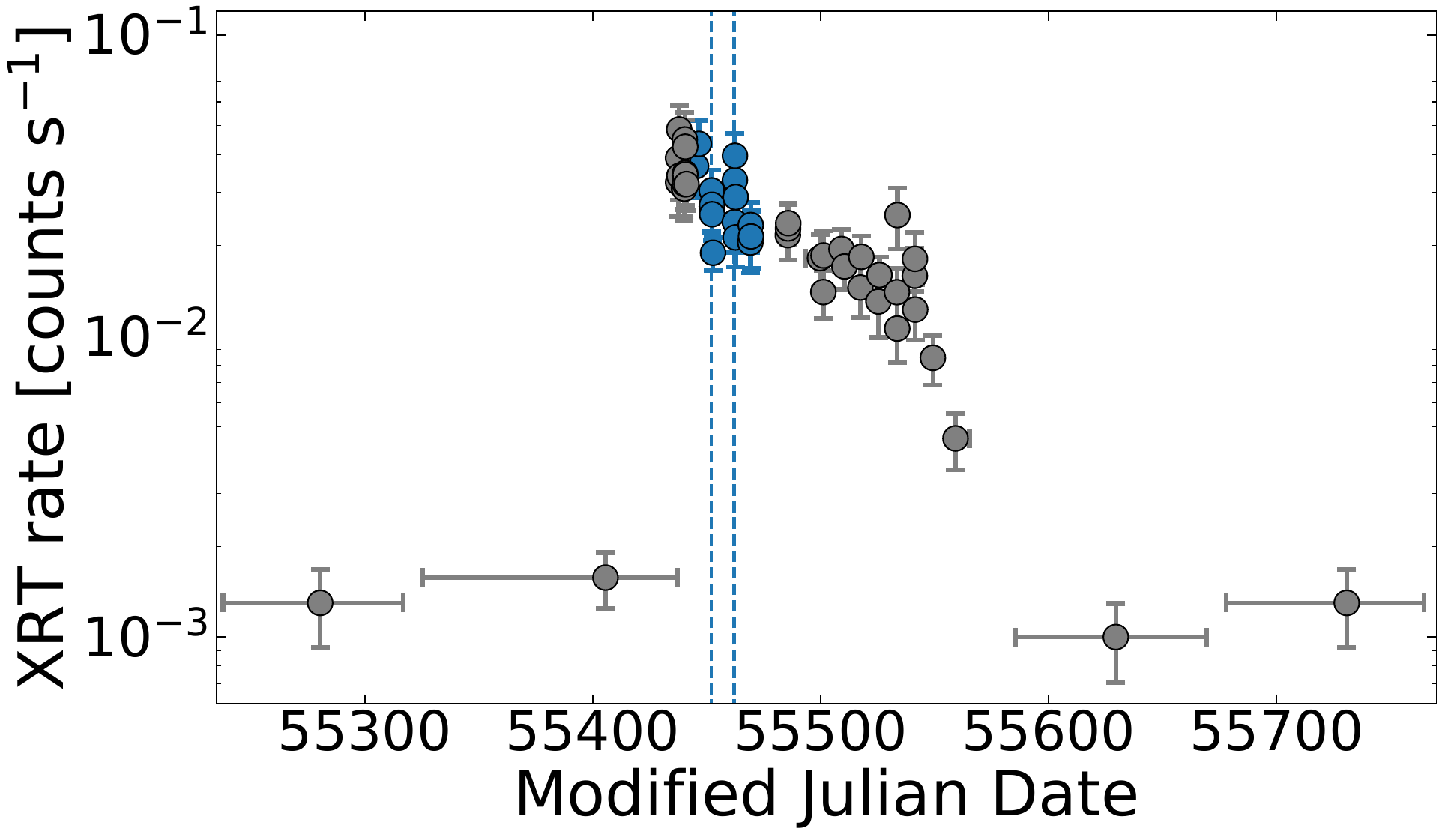}
	\caption{X-ray count rate light curve of the 2010 outburst of \hlx. Dashed vertical lines mark the epochs of the \hst observations, and the blue point represents the observations ($\pm 10$ days from \hst observations) that were stacked to create \hlx's X-ray spectrum in the soft/high state.}
    \label{fig:lc_hlx}
\end{figure}

Many authors have pointed out that the ``missing energy'' problem is ``solved'' by: i) most of the energy being released in the EUV \citep{Dai2018,Lu18,Thomsen2022_disk_wind, Mummery2023,Guolo2024}; and ii) most of the energy being released at time scales much longer ($\gg 1$ yr) than the initial flare \citep{Mummery2021b}.Our analyses of \ftli presented here agrees with both, as the X-ray and the UV/optical band carry just a small fraction of the total energy, and \ftli's $L_{\rm Bol}$ is $\sim 1\times 10^{44}$ \ergs (or $\lambda_{\rm Edd} \sim 0.2$, for the most likely \Mbh value) almost four years after disruption, as shown by Fig.~\ref{fig:14li_sed} and Fig.~\ref{fig:BB_comp}. These findings emphasize two key conclusions: i) the bolometric luminosity and/or the accretion rate do not follow the fallback rate ($\propto t^{-5/3}$) but instead evolves on much longer timescales, as long predicted by time-dependent disk theory \citep[e.g.,][]{Cannizzo1990,Mummery2020}; ii) there is no energy missing from \ftli.

\subsection{HLX-1}\label{sec:HLX}

\hlx is an off-nuclear variable X-ray source in the nearby ($z = 0.0223$) edge-on spiral galaxy ESO243-49 \citep{Farrell2009}. Its maximum 0.2-10 keV luminosity of up to $\sim 1 \times 10^{42}$ \ergs makes a lower black hole mass (\Mbh $\leq$ 500 \msun) very unlikely, positioning the source as one of the best candidates for the elusive class of intermediate-mass black holes \cite[IMBH, see][for a review on IMBHs]{Greene2020}. Similar to lower luminosity X-ray binaries (XRBs) and ultra-luminous X-ray sources (ULXs), \hlx has exhibited multiple outbursts, transitioning between hard/low and soft/high spectral states \citep[][and references therein]{Soria2017}, where the X-ray spectrum shifts from a power-law to a thermal shape. A UV/optical/IR counterpart has long been identified \citep{Soria2010}, but its physical origin has been the subject of intense debate \citep{Soria2010,Farrell2012,Farrell2014,Webb2014,Soria2017}, with interpretations varying between distinct combinations of direct disk emission, reprocessed disk emission, and young and/or old stellar populations. However, the factor of a few variability in all bands (from FUV to NIR) during the X-ray outbursts \citep[see Figure 4 of][]{Soria2017} makes the dominance of a stellar population quite unlikely, suggesting a disk-related origin is much more probable.

Our model implementations, as described in \S\ref{sec:model}, should be able to shed light on this problem. If the model accurately describes the data in the soft/high state, it should result in physically meaningfully values for the system's parameters. For our broad-band spectrum analyses, we combine the \hst data, as described in \S\ref{sec:data}, with a \swift/XRT spectrum resulting from stacked observations taken within $\pm 10$ days around the \hst observations during the soft/high state of the 2010 outburst, as shown in the light curve in Fig~\ref{fig:lc_hlx}.

We start our analyses in the Newtonian regime and simlar to the previous section apply the model \texttt{phabs$\times$redden$\times$\texttt{zashift}(phabs$\times$reddenSF$\times$diskSED)}. The Galactic X-ray neutral gas absorption is given by the fixed $N_{H, G} = 2.0 \times 10^{20}$ cm$^{-2}$, and the Galactic extinction by E(B-V)$_{G}$ = 0.021. The intrinsic part of the model is shifted to the source rest frame using $z = 0.0223$. For the same reasons as discussed in \S\ref{sec:14li}, we linked the intrinsic neutral gas X-ray absorption and intrinsic UV/optical dust attenuation by a Galactic-like gas-to-dust ratio \citep{Guver2009}. Uniform priors are assumed for the four free parameters.

\begin{figure*}[t]
	\centering
	\includegraphics[width=1.0\textwidth]{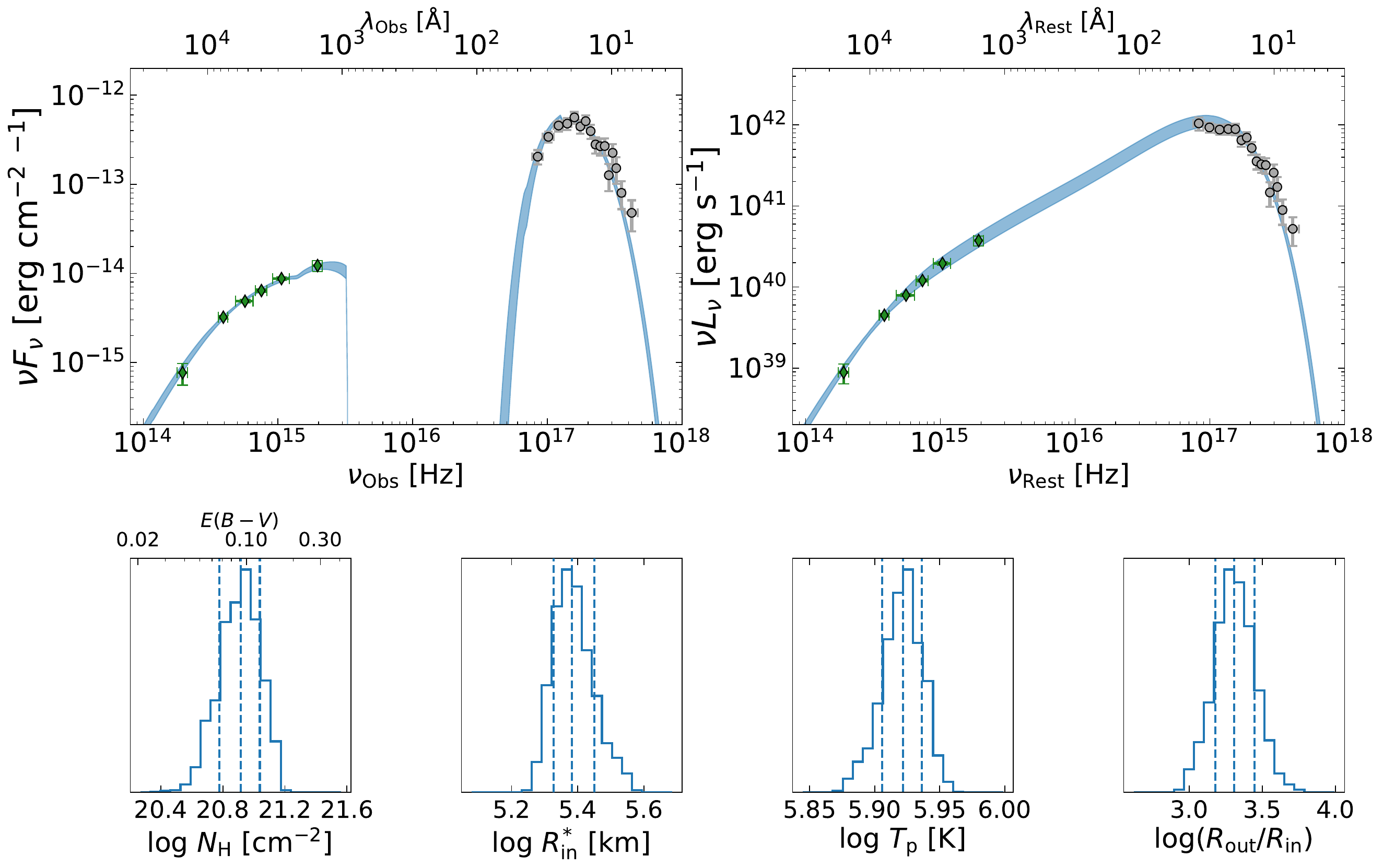}
	\caption{Results of the nested sampling fit of \texttt{diskSED} to broad-band data of \hlx. Panels are the same as in Fig.~\ref{fig:14li_sed}.}
    \label{fig:hlx_sed}
\end{figure*}

The results of the nested sampling fit are shown in Fig.~\ref{fig:hlx_sed}. The bottom panel displays the 1D projection of the four parameter posteriors, with the full posterior in Appendix \S\ref{sec:app_post}. The convergence of the sampling is clear. In the upper left panel, we show the observed flux models (without any extinction/absorption corrections) overlaid on the observed UV/optical photometry and the unfolded X-ray spectrum. The right panel shows the intrinsic luminosities (with both Galactic and host-galaxy attenuation and absorption corrections), with data points unfolded to the median values of the parameter posteriors. 

The extended nature of the disk is evident (unlike what was observed for \ftli) from the extremely long mid-frequency $\nu L_{\nu} \propto \nu^{4/3}$ portion of the broad-band spectrum and the transition to the Rayleigh-Jeans regime occurring only in the optical red/IR bands. Higher \Tp and lower \Rins values, as expected from \hlx's presumed IMBH nature are obtained.

\begin{figure}[b]
	\centering
	\includegraphics[width=0.9\columnwidth]{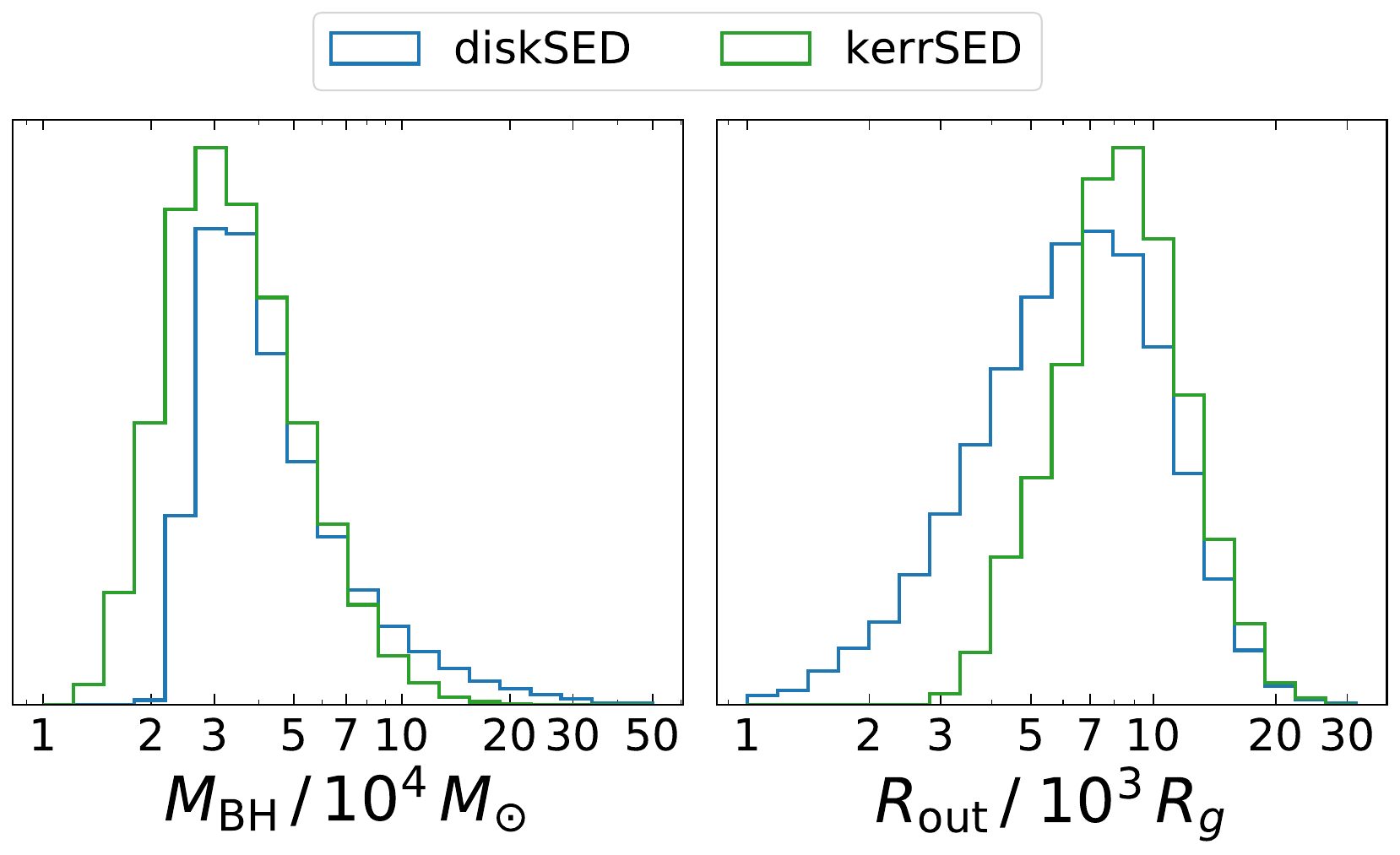}
	\caption{Probability distribution functions for \hlx's \Mbh (left panel) and outer disk radius ($R_{out}$) in gravitational radius's ($R_g$, right panel). Blue distribution for \texttt{diskSED} fit and green for \texttt{kerrSED} fit.}
    \label{fig:HLX1_MBH}
\end{figure}

\begin{figure}[b]
	\centering
	\includegraphics[width=0.8\columnwidth]{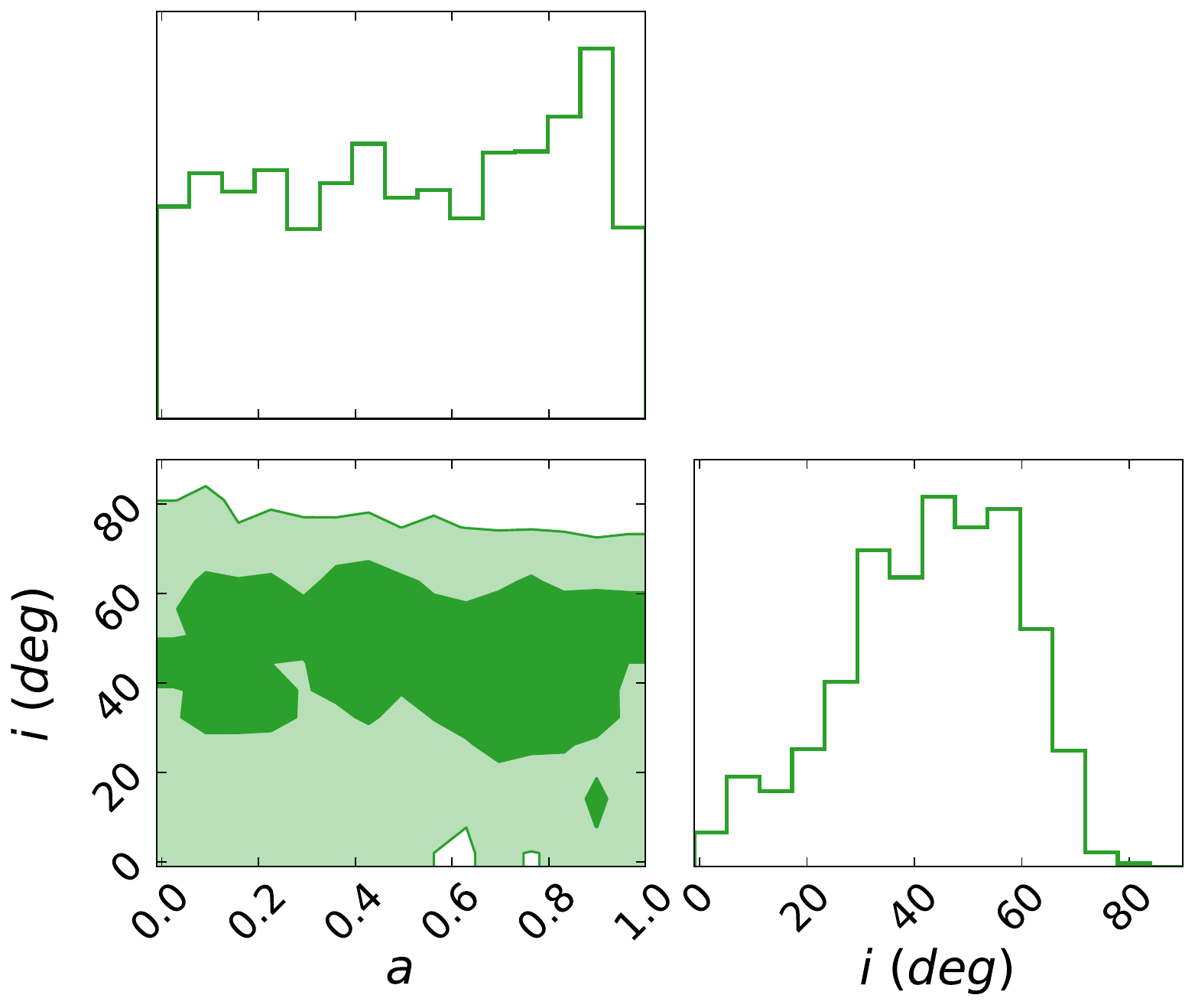}
	\caption{Projection of the posterior distribution for inclination ($i$) and spin ($a$) for the \texttt{kerrSED} fit of HLX-1. In the 2D histogram, contours represent 68\% and 99\% of the distribution. The full posterior, including the remaining free parameters, is shown in Appendix \S\ref{sec:app_post}.}
    \label{fig:a_x_i_HLX1}
\end{figure}

Similarly to discuss in the previous section, we can infer \hlx's \Mbh from the \Rins, under assumptions about ${\rm cos}\, i$ and $a$. 
Similar to \ftli, we simple assume a flat distribution of possible spins in the range $0 \leq a \leq 0.99$. For the inclination, there are no independent (of X-ray continuum fitting) estimates, and we simple assume flat probability distribution of ${\rm cos}\, i$, with inclinations in the full range $0^\circ \leq i \leq 85^\circ$. The probability distribution of \Mbh for \hlx is shown in blue in left panel of Fig.~\ref{fig:HLX1_MBH}, can be written as $M_{\rm BH} = 5^{+8}_{-2} \times 10^{4} M_{\odot}$, supporting the IMBH nature of the source. Under the same uninformative spin distribution assumption, an $R_{\rm out} \approx {\rm few} \times 10^3 R_g$ is obtained, which indicates an extremely old accretion system and/or a disk fed by a wide binary, and is similar to values estimated from XRB and ULXs \citep[e.g.,][]{Remillard2006}. 

Our relatively high uncertainty on \Mbh, particularly the high end skewing of the probability distribution is mainly driven by our completely ignorance on the inclination of the system, and its influence on the \Mbh value (see Eq. \ref{eq:M_BH}). This motivates us to try to obtain some constraint on the completely unknown values of $a $ and $i$, using \texttt{kerrSED}. We apply the model \texttt{phabs$\times$redden$\times$zashift} 
\texttt{(phabs$\times$reddenSF$\times$kerrSED)} to the same data, using the same 
values/constraints and flat priors for the other parameters allowing $i$ and $a$ to 
vary freely, and assuming flat prior for these as well. The full parameters posterior 
is shown in Appendix \S\ref{sec:app_post}, and in Fig.~\ref{fig:a_x_i_HLX1} we show the 2D 
projection of $a \times i$ plane of the posterior, alongside the 1D projection of the 
two parameters posterior. As one would expect, and as discussed in 
\S\ref{sec:model_kerrSED}, no information can be obtained from the spin ($a$), given its
subtle effects and relatively low S/N of the X-rays spectrum. However, some information 
can be inferred about the inclination, as the model seems to be able to completely 
exclude edge-on configurations, slightly disfavors face-on configurations, and has most
of its posterior mass equally distributed in the range $10^\circ \leq i \leq 70^\circ$.
As a sanity check, we see that the recovered values of the remaining parameters are consistent with those from \texttt{diskSED}. A slight increase in $T_p$ ($\sim 0.05 {\rm dex}$) is attributed to the gravitational redshift effects on the X-ray photons propagating through the Kerr metric, requiring a small increase in $T_p$ to produce the same X-ray flux. With \texttt{kerrSED}'s results we can now infer \Mbh and 
$R_{\rm out}/R_g$ using the posterior values of $a$ and $i$ instead of flat ad hoc 
distribution. As shown in green in Fig.~\ref{fig:HLX1_MBH}, the \Mbh distribution is 
narrow, hence the inferred \Mbh are more concentrated at values that can be described as $M_{\rm BH} = 4^{+3}_{-1} \times 10^{4} M_{\odot}$, a slight
improvement on $R_{\rm out}/R_g$ was also obtained, but the values is still consistently at 
$R_{\rm out} \approx {\rm few} \times 10^3 R_g$.

From our full SED fitting, the Bolometric luminosity is easily estimated by integrating under the model (values from \texttt{diskSED} and \texttt{kerrSED} are consistent), resulting in $L_{\rm Bol} = 1.8\pm0.1 \times 10^{42}$ \ergs. For the same epoch, the Eddington ratio ($\lambda_{\rm Edd} = L_{\rm Bol}/L_{\rm Edd})$ is therefore $0.15\pm0.015$ (assuming \texttt{kerrSED}'s \Mbh), given the analyzed epoch is slight fainter than the peak of the outburst (see Fig.~\ref{fig:lc_hlx}) this means that \hlx reaches $\lambda_{\rm Edd} \lesssim 0.25$ at its outburst peak. 

The values obtained here for \Mbh and $R_{\rm out}$, are in agreement to the first order, and given uncertainties and distinct assumptions, with several other estimates of these two values by many other authors \citep[e.g.,][]{Servillat2011,Davis2011,Godet2012,Straub2014,Webb2014,Soria2017}. It is important to notice, however, that most of these multi-wavelength analyses of \hlx had employed much more complex models, e.g. the disk emission was usually modeled using \texttt{diskir} \citep{Gierlinski2008}, which employ a series of additional effects (therefore added free parameter), which from our fitting are not clearly necessary. As an example, \citet{Soria2017}'s modeling
of the same soft/high state, had between 8 and 11 total free parameters. Detailed statistical model comparison is beyond the scope of this paper, but an increase from our 4 (or 6 in the relativistic case) to 8-11 free parameters (none related to GR corrections) seems unlikely to be justified given the results of Fig.~\ref{fig:HLX1_MBH} and Appendix \S\ref{sec:app_post}. We however support the conclusion of the authors that the UV/optical emission from \hlx is dominated by accretion not from a young stellar population. Speculations about the origin of the accretion material, or the mechanism behind the outburst and state transitions in \hlx are beyond the scope of this spectral modeling paper.

\section{Notes on Model usage and limitations}\label{sec:notes}

In this section we discuss some aspect of the applicability of the models presented in this work and some of its limitations.

\textbf{Newtonian vs. Relativistic:} The relativistic case adds two free parameters compared to the Newtonian case. In a frequentist framework, this may not be justified for X-ray spectra of black holes in the mass range of interest due to the limited counts available. However, in a Bayesian framework,  these ``nuisance'' parameters  can be marginalized over, such that even if their posterior do not converge completely, at least some regions in the $a \times i$ plane of the parameter space may be excluded. By narrowing down the parameter space, we can derive more precise inferences for other physical quantities, such as \Mbh, as we have demonstrate in \S\ref{sec:HLX}. It should be noted, however, that the numerical ray tracing in \texttt{kerrSED} makes the model significantly ($\gg 10\times$) slower than \texttt{diskSED}, requiring high computational power for extended parameter space sampling.  Further, given the much subtle effects of Relativistic corrections, and the count rate regime of X-ray spectra of  sources in the space parameter of interest, the authors advice that users use \texttt{kerrSED} with Bayesian inference methods (e.g., MCMC, nested sampling, etc) and do not recommend the usage of the model in a frequentist framework, e.g., the native Levenberg–Marquardt minimizer in \texttt{XSPEC} \citep[see][for some statistical discussion]{Andrae2010,Buchner2023}. Therefore, the usage of a classical or relativist disk model, needs to be consider on a case-by-case manner, depending on the goals of the research and the quality and wavelength coverage of the data.

\textbf{Conditions for application to TDEs:} The early-time optical/UV emission, which is typically used to discover optically-selected TDEs, is known not to originate from direct disk emission. As a result, applying the models presented in this work during these early phases ($\leq 1$ yr) is unlikely to yield convergent fits and may produce unphysical parameters or evolution trends. The hypothesis proposed here, which has been validated for ASASSN-14li but remains to be tested on a larger sample, is that these models become applicable once the UV/optical emission transitions from a rapidly decaying component to a more slowly evolving `plateau' phase. Users aiming to apply these models might follow the approach of \citet{Mummery2024}, fitting the UV/optical light curves with a combination of a power-law (or exponential) decay and a constant (or linear) component. It is crucial to ensure that the latter component dominates the emission in the epochs when the full SED fitting is aimed to be performed. It is important to emphasize that while the temperature profile described by Eq.~\ref{eq:T_r} can be formally derived under the assumption of a steady-state disk, such a profile also arises asymptotically in time-dependent TDE disk solutions \citep{Cannizzo1990, Mummery2020}. This occurs provided that the stress vanishes at the ISCO (a key assumption in this model) and that the disk has evolved beyond the bolometric luminosity's rising phase. More specifically, this condition is satisfied when the disk has undergone at least one evolutionary timescale, i.e., $t \gtrsim t_{\rm evol}$ \citep{Mummery2020,mummery2024fitted}. While in general this should be satisfied during the `plateau' phase, it is not necessarily the case that the early optical/UV emission in a TDE evolves on the same timescale as the disk (this would explicitly not be the case for shock-powered emission, for example). Some care should be taken in interpreting results of this analysis during the rising phase of TDE X-ray lightcurves, even if the optical/UV luminosity has plateaued. 

\textbf{Inclusion of a Hard X-ray Component:} The two examples presented here feature sources in a fully soft/thermal state. However, this may not always be the case, as a hard/Comptonized component can also be present. When both thermal and hard components are observed i.e., the source is in an `intermediate state', the models discussed in this work can still be applied by incorporating a physically motivated Comptonization model, such as \texttt{simPL} \citep{Steiner2009}, into the fitting process. If only spectra in a hard (power-law dominated) state are available, the model should not be used because the underlying inner disk properties cannot be uniquely determined from a purely coronal spectrum (see, for instance, the simulations in Appendix A of \citet{Guolo2024} for TDE cases). However, if spectra from both hard and soft states are available, it might be possible to derive some disk properties of the hard state by imposing physically motivated priors based on the soft state results. Nonetheless, the results in such case should be carefully and critically interpreted.

\begin{figure}[h]
	\centering
	\includegraphics[width=1.0\columnwidth]{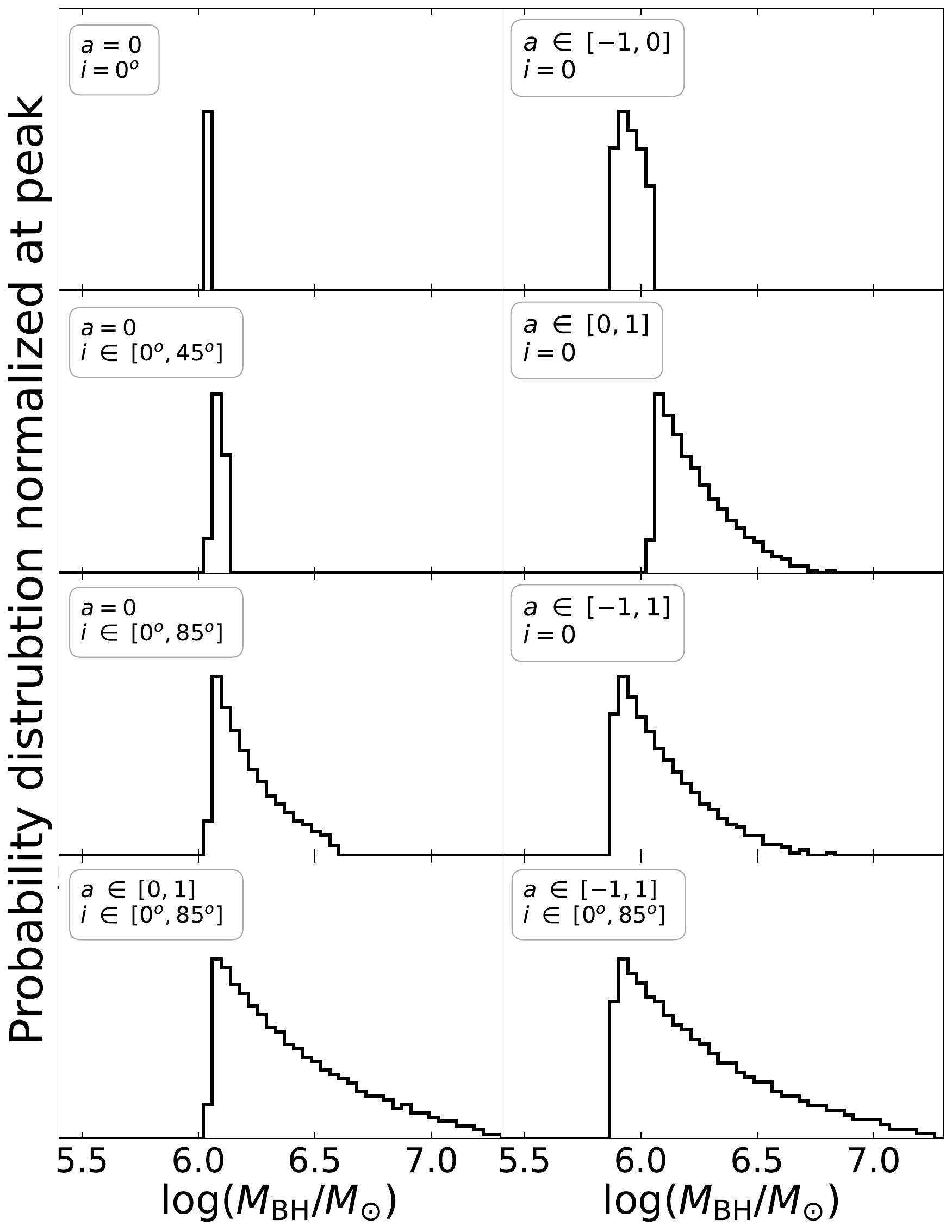}

	\caption{Systematics effects of spin ($a$) and inclination ($i$) assumptions on the recovered \Mbh using \texttt{diskSED}. The panels whow the probability distribution functions of the resulting \Mbh, for a fixed \Rins $= 10^7$ km, under distinct assumption of $a$ and $i$. When a range is quoted, a flat distribution in the range is assumed, in the case of $i$ the assumed distribution is flat on ${\rm cos} \ i$ space.}
    \label{fig:MBH_syst}
\end{figure}

\textbf{Limitations of \Mbh estimates on the Newtonian limit: }  In Eq. \ref{eq:M_BH}, \Mbh is determined by three variables; 
however, in the Newtonian limit, only one of these (\Rins) can be directly constrained by the data. As a result, \Mbh estimates and their uncertainties, obtained with \texttt{diskSED} will be naturally dependent on the \textit{ad hoc} assumptions made about spin and inclination. Specifically, \Mbh is inversely proportional to both $\sqrt{\cos \ i}$ and $\gamma(a)$ (as defined in Eq. \ref{eq:gamma}). This implies that even if the spin and inclination are assume to be uniform distributed, the resulting distribution of \Mbh can be highly asymmetric due to the nonlinear dependence on these parameters. In most cases the \Mbh uncertainties will be dominated by these systematic effects instead of data/statistical uncertainties.  To demonstrate these effects, we compute the resulting \Mbh probability distribution functions for a fixed \Rins = $10^7$ km, considering different assumptions and distributions for $a$ and $i$. The resulting distributions of \Mbh are shown in the bottom panels of Fig.~\ref{fig:MBH_syst}. For the $R_{\rm out}/R_g$ estimates, the distributions will also be dependent on the assumptions, but only on the spin one, and proportional to $\gamma(a)$ instead of 1/$\gamma(a)$.

\textbf{Data requirements and parameter inference effects:} A key result of this work is that simultaneous multi-wavelength fitting yields more physically meaningful results than single-band fitting. However, these models are still X-ray spectral fitting models, hence a soft X-ray spectrum with at least a few energy bins above the background is essential. This is because accurately determining \Tp is crucial for inferring other key parameters like \Rratio and \Mbh.
For lower-energy data, the constraining power and the data requirements for achieving full convergence versus one-side convergence, or no convergency at all, depend not only on the physical properties of the underlying system but also on the specific wavelength coverage available; in a non-trivial manner. Although some intuition can be obtained by investigating the asymptotic limit of the equations describing the model and Fig.~\ref{fig:model_pars} and \ref{fig:model_pars_kerr}. One advantage of employing a Bayesian framework is that the constraining power of the available data is directly reflected in the marginalized posterior distributions, as per the Bayes' theorem. Preliminary applications show varying constraint powers; a single UV photometric point can fully constrain \Rratio if the coverage reaches the Rayleigh–Jeans tail \citep{Wevers2025}, while in other cases, even an entire UV spectrum may only yield a lower limit on  \Rratio, leading to a one-side converged posterior \citep{Guolo2025}, and needing an optical or IR observation to fully converge.

\section{Conclusions}\label{sec:conclusion}
In this paper, we have implemented two models tailored for simultaneous and self-consistent fitting of X-ray spectra and UV/optical/NIR photometric data of accreting black holes in a thin disk state. These models are integrated into the standard X-ray fitting package, \texttt{pyXspec}. We demonstrated the application of these models by fitting the multi-wavelength emission of two distinct systems: the TDE \ftli in its late-time ``plateau" phase, and the IMBH candidate \hlx in its soft/high state.

\noindent Regarding the implemented models:

\begin{itemize}
\item In the Newtonian limit, \texttt{diskSED} describes the broad-band spectrum of a standard thin disk with a well-defined ratio between the outer and inner radii (\Rratio) and a physical peak disk temperature (\Tp). The model normalization is given by the parameter \Rins ($=R_{\rm in} \sqrt{{\rm cos}\,i}$). The black hole mass (\Mbh) can be inferred from \Rins under assumptions about the inclination ($i$) and spin ($a$).
\item In the relativistic regime, \texttt{kerrSED} describes a standard thin disk in the Kerr metric by including numerical ray tracing calculations of the photon's propagation. The inclination ($i$) and the spin ($a$) are the two additional free parameters that can be marginalized over as part of the fitting.
\end{itemize}

\noindent For the application to \ftli, we fit three epochs in the ``plateau'' phase, from approximately 350 days to 1300 days after discovery using \texttt{diskSED}. Our conclusions are as follows:

\begin{itemize}
\item We show that at these late times, the multi-wavelength emission of the TDE can be fully described by a standard thin disk.
\item We obtain log(\Rins/km) = 7.6-7.7, consistently between the three epochs, which, under reasonable assumptions about $a$ and $i$, results in an inferred $M_{\rm BH} = 7^{+3}_{-2}\times10^{6} M{\odot}$, in agreement with many other estimates.
\item The predicted cooling of the disk is recovered with high significance.
\item A compact disk, with $R_{\rm out}$ of $45 \pm 13 \, R_{\rm g}$ -- consistent with the circularization radius -- is obtained at the first epochs. There is possible expansion at the third epoch to $R_{\rm out} \leq 120 , R_g$ (99\% posterior), though this outer radius is still statistically consistent with the results of the first epoch.
\item The standard $L_{\rm Bol} \propto T_p^4$ relation describes well the evolution of the bolometric  emission, but the X-ray luminosity has a much steeper dependence on temperature, as expected.
\item The total energy emitted from $\Delta t = 350$ to $\Delta t = 1250$ was $\sim 2 \times 10^{52}$ ergs (or $\sim 0.1$ \msun, assuming 10\% efficiency), with most energy emitted in the EUV. The source is still emitting $L_{\rm Bol} \approx 10^{44}$ \ergs at $\sim$ 3.5 years after disruption.
\item We discuss at length the advantages of our modeling over simplistic single-temperature blackbody fits, in which X-ray and UV/optical data are independently fitted.
\end{itemize}

\noindent Regarding the model fitting for the high/soft state of \hlx:

\begin{itemize}
\item We show that the multi-wavelength emission from X-ray to NIR can be described by a thin disk without the need for any additional stellar population component.
\item Higher \Tp and lower \Rins (compared to \ftli) are obtained, consistent with a lower \Mbh.
\item An extremely extended disk, with $R_{\rm out} \approx {\rm few} \times 10^{3} \, R_g$, is recovered -- given that the transition from the mid-frequency range ($\nu L_{\nu} \propto \nu^{4/3}$) to the Rayleigh-Jeans tail occurs only at the red optical to NIR bands, indicating a long-lived accretion flow and/or fed by a wide binary.
\item By fitting the \texttt{kerrSED} model, we show that intermediate inclinations of $10^\circ \leq i \leq 70^\circ$ are preferred over either face-on or edge-on configurations. However, no constraint on the spin ($a$) can be obtained, given the only moderate S/N of the X-ray spectrum.
\item The \texttt{kerrSED} fit results in a well-constrained black hole mass of $M_{\rm BH} = 4^{+3}_{-1} \times 10^4$ \msun, in agreement with previous studies and consistent with the IMBH nature of \hlx.
\end{itemize}
%\begin{acknowledgements}

\textit{Acknowledgements} -- 
MG is grateful to S. Gezari, T. Wevers, and Y. Ajay for fruitful discussion about this work, and for providing comments on the early versions of the manuscript, specially thanks to Y. Ajay for providing us the reduced \xmm data of \ftli. MG is partially supported by NASA \nicer grant 80NSSC24K1203. This work was supported by a Leverhulme Trust International Professorship grant [number LIP-202-014]. 
This work made use of data supplied by the UK Swift
Science Data Centre at the University of Leicester.

\facilities{HST, Swift, XMM}

\software{ {\tt matplotlib} \citep{Hunter2007}, {\tt scipy} \citep{scipy}, {\tt numpy} \citep{numpy}, {\tt astropy} \citep{astropy},  {\tt XSPEC} \citep{Arnaud1996}, {\tt BXA} \citep{Buchner2014}, {\tt UltraNest} \citep{Buchner2019}, {\tt corner} \citep{corner}, .}
\appendix

\section{Marginalized Posteriors}\label{sec:app_post}

\begin{figure*}[h]
	\centering
	\includegraphics[width=0.8\columnwidth]{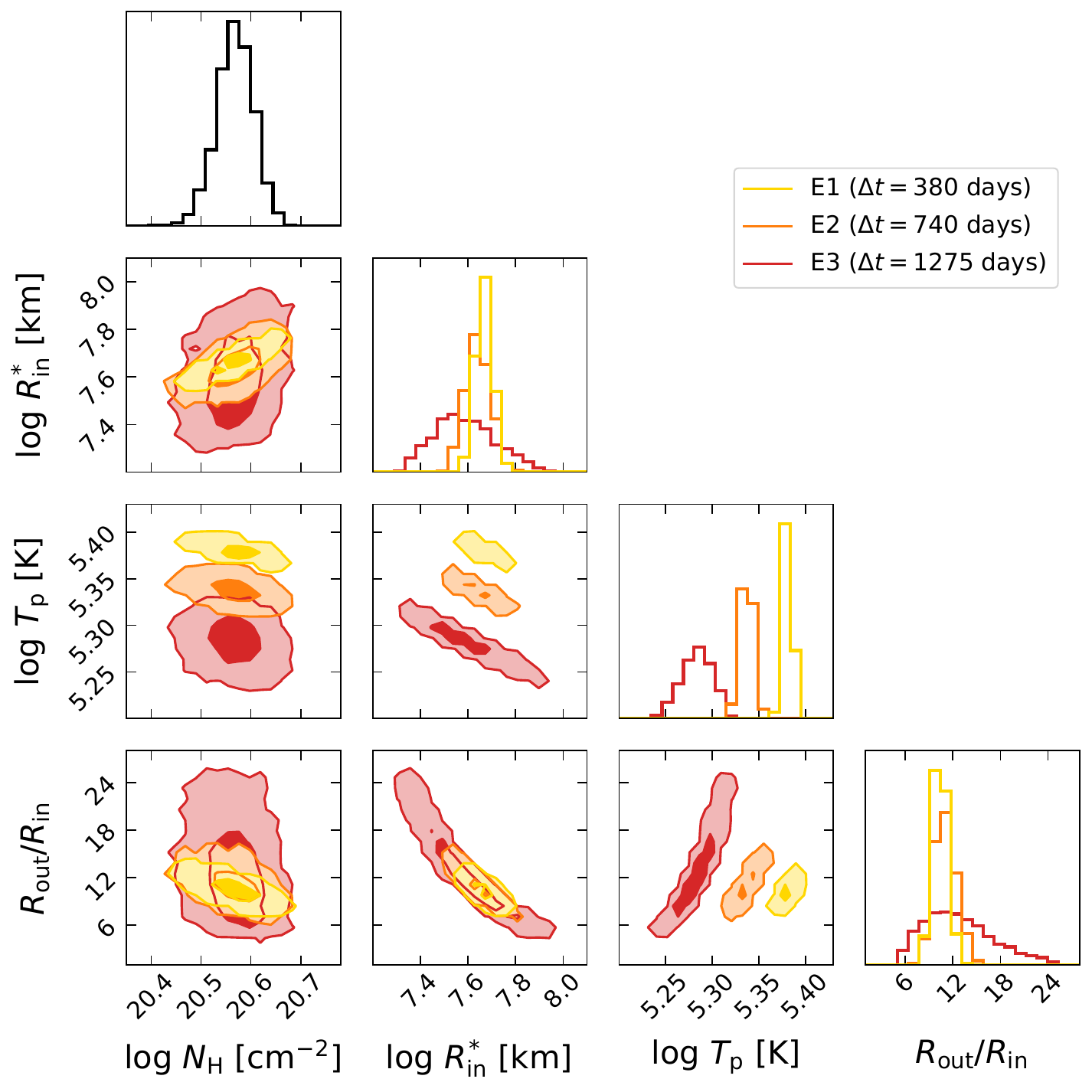}
	\caption{Full Marginalized posterior for the \texttt{diskSED} fit to the three epochs of \ftli: E1 (Yellow), E2 (orange), and E3 (red). In the 2D histogram the contours shows 68\% and 99\% of the probability distribution.}
    \label{fig:14li_post}
\end{figure*}

\begin{figure*}[h]
	\centering
	\includegraphics[width=0.8\columnwidth]{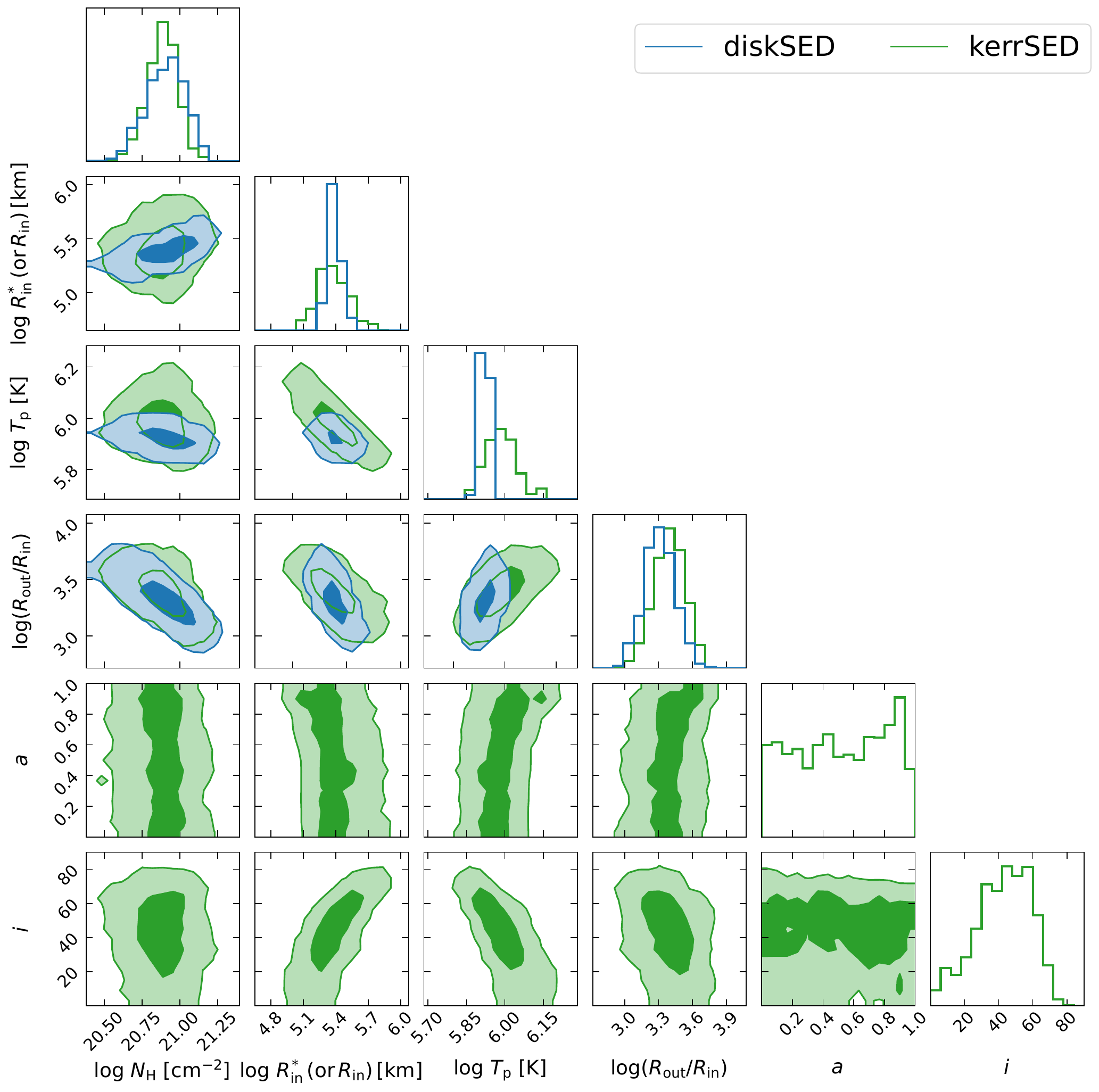}
	\caption{Full marginalized posterior for the \texttt{diskSED} (blue) and 
 \texttt{kerrSED} (green) fit to the soft/high state of \hlx. For \texttt{diskSED} the four free parameters are $N_{\rm H}$, \Rins, \Tp, and \rm{log}(\Rratio); while for \texttt{kerrSED} the six free parameters are $N_{\rm H}$, \Rin, \Tp, \rm{log}(\Rratio), $a$, and $i$. In the 2D histogram the contours shows 68\% and 99\% of the probability distribution. }
    \label{fig:HLX_post}
\end{figure*}
\bibliography{tde}{}
\bibliographystyle{aasjournal}

\end{document}